\documentclass[journal,letterpaper]{IEEEtran}

\usepackage{amsmath}
\usepackage{amssymb}
\usepackage{amsfonts}
\usepackage{graphicx}
\usepackage{epsfig}
\usepackage{subfigure}
\usepackage{psfrag}
\usepackage{cite}
\usepackage{color}

\usepackage[letterpaper,margin=0.8in]{geometry}
\usepackage{dblfloatfix}

\begin{document}
\title{Massive Connectivity with Massive MIMO--Part I: Device Activity Detection and Channel Estimation} 
\author{Liang Liu, \IEEEmembership{Member,~IEEE} and Wei Yu, \IEEEmembership{Fellow,~IEEE}

\thanks{Manuscript received June 17, 2017, revised November 9, 2017 and January 30, 2018, accepted March 8, 2018. The materials in this paper have been
presented in part at the IEEE International Symposium on Information Theory
(ISIT), Aachen, Germany, July 2017 \cite{LiangISIT}. The associate editor coordinating the review of this paper and approving it for
publication was Dr. Mathini Sellathurai. This work is supported by Natural Sciences and Engineering
Research Council (NSERC) of Canada via a discovery grant, a Steacie Memorial Fellowship
and the Canada Research Chairs program.}
\thanks{The authors are with The Edward S. Rogers
Sr. Department of Electrical and Computer Engineering, University of Toronto,
Toronto, Ontario, Canada, M5S3G4,
(e-mails:lianguot.liu@utoronto.ca; weiyu@comm.utoronto.ca).}}

\maketitle

\begin{abstract}
This two-part paper considers an uplink massive device communication
scenario in which a large number of devices are connected to a
base-station (BS), but user traffic is sporadic so that in any given
coherence interval, only a subset of users are active. The objective
is to quantify the cost of active user detection and channel
estimation and to characterize the overall achievable rate of a
grant-free two-phase access scheme in which device activity detection
and channel estimation are performed jointly using pilot sequences
in the first phase and data is
transmitted in the second phase.  In order to accommodate a large
number of simultaneously transmitting devices, this paper studies an
asymptotic regime where the BS is equipped with a massive number of
antennas.  The main contributions of Part I of this paper are as
follows.  First, we note that as a consequence of having a large pool
of potentially active devices but limited coherence time, the
pilot sequences cannot all be orthogonal.  However, despite the
non-orthogonality, this paper shows that in the asymptotic massive
multiple-input multiple-output (MIMO) regime, both the missed device
detection and the false alarm probabilities for activity detection can
always be made to go to zero by utilizing compressed sensing
techniques that exploit sparsity in the user activity pattern.
Part II of this paper further characterizes the achievable rates using
the proposed scheme and quantifies the cost of using non-orthogonal
pilot sequences for channel estimation in achievable rates.
\end{abstract}


\IEEEpeerreviewmaketitle

\newtheorem{corollary}{Corollary}
\newtheorem{definition}{Definition}
\newtheorem{lemma}{Lemma}
\newtheorem{theorem}{Theorem}
\newtheorem{proposition}{Proposition}
\newtheorem{remark}{Remark}
\newcommand{\mv}[1]{\mbox{\boldmath{$ #1 $}}}

\section{Introduction}\label{sec:Introduction}

\subsection{Motivation}

Massive connectivity is a key requirement for future
wireless cellular networks that aim to support Internet-of-Things
(IoT) and machine-type communications (MTC).
In a massive device connectivity scenario, a cellular base-station
(BS) may be required to connect to a large number of devices (in the
order $10^4$ to $10^6$), but a key characteristic of the IoT and MTC
traffic is that the device activity patterns are typically
\emph{sporadic} so that at any given time only a small 
fraction of potential devices are active \cite{Bockelmann}. The sporadic
traffic pattern may be due, for example, to the fact that often
devices are designed to sleep most of the time in order to save energy
and are activated only when triggered by external events, as typically
the case in a sensor network.
In these scenarios, the BS needs to dynamically identify the active
users before data transmissions take place. 
The central question this paper seeks to address is how to quantify
the cost of device activity detection and to account for its
impact on the overall achievable rate in the cellular system design.

\begin{figure}
\begin{center}
\scalebox{0.5}{\includegraphics*{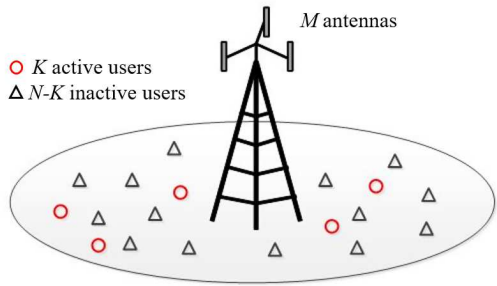}}
\end{center}
\caption{System model of the massive device communication network.}\label{fig1}
\end{figure}

\begin{figure}
\begin{center}
\scalebox{0.4}{\includegraphics*{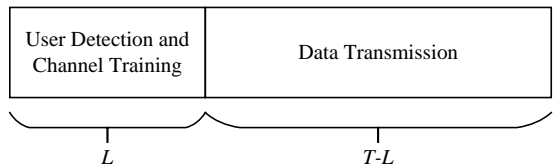}}
\end{center}
\caption{Two-phase transmission protocol.}\label{fig2}
\end{figure}

This paper considers the uplink of a single-cell massive connectivity
scenario consisting of a BS equipped with $M$ antennas connected to
a pool of $N$ potential devices, of which a fraction of $K$ devices are active at
any given time, as shown in Fig.~\ref{fig1}. In each coherence time interval $T$, the BS needs to
identify the active devices, to estimate their channels and to decode
the transmitted messages from the devices. In particular, a two-phase
multiple-access scheme is adopted in which joint activity detection and
channel estimation are performed using pilot sequences in the first
phase of duration $L$, while data transmission takes place in the second
phase of duration $T-L$, as shown in Fig.~\ref{fig2}. Note that we choose to combine user activity
detection and channel estimation in order to minimize scheduling latency.

Observe that in order to achieve an appreciable data rate, $M$ has
to be in the same order as $K$. This intuition can be justified
from a degree-of-freedom (DoF) perspective by noting that when
coherence time is large, the achievable DoF of a multiple-access
system (without assuming a priori channel state information (CSI) at
the receiver) is at maximal when the number of active devices equals to
the number of receive antennas at the BS \cite{yu_ITA}.
To accommodate a large number of active devices, the above result
motivates us to consider the massive multiple-input multiple-output
(MIMO) regime with large $M$.

Further, in order to achieve reasonably accurate uplink channel estimation,
$L$ needs to be larger than $K$. Yet, since $L$ is constrained by
the coherence time and the total number of potential devices $N$ can be
very large, we usually have the situation where $N>L$. Thus, it is
typically not possible to assign orthogonal pilot sequences to all the
potential devices.

The main objectives of this paper are to quantify the performance of
device activity detection and channel estimation when \emph{randomly
generated non-orthogonal pilot sequences} are assigned for each
device, and to examine its impact on the overall achievable data rate.
Toward this end, we propose the use of approximate message passing
(AMP) algorithm \cite{donoho_amp} in compressed sensing to exploit sparsity in device
activity detection. We further analyze the resulting device detection
performance and the channel estimation error by utilizing large-system
result where $N$, $K$, $M$ and $L$ go to infinity in certain limits,
thereby allowing a characterization of the overall achievable data
rate for massive connectivity.

The main theoretical results of Part I of this paper show that massive
MIMO is especially well suited for massive device connectivity in the
sense that device activity detection error can always be driven to zero
asymptotically in the limit as $M$ goes to infinity. 
Despite the asymptotically perfect detection, however, this paper also
shows that the channel estimation error always remains---because of
the use of non-orthogonal pilots---and that channel estimation error
is the main cost of user activity detection on achievable rate. The
analytical results of Part II of this paper \cite{LiuPart2}
quantify this additional channel estimation error and shows its impact on
achievable rate and optimal pilot length for massive connectivity.

\subsection{Prior Work}

Conventional cellular networks are designed based on the scheduling of
active users to time or frequency slots. The overhead of scheduling
massive number of sporadically active users over a separate control
channel can, however, incur significant overhead.
To this end, contention-based schemes have been proposed to deal with
this issue.  For example, \cite{Niyato,Pratas,Bjornson} investigate
a random access protocol in which each active user picks one of the
orthogonal signature sequences at random and sends it to the BS, and
a connection is established if the selected preamble is not used by
the other users. The main drawback of this random access protocol is
the need for contention resolution, since collision is unavoidable
with a massive number of devices.

As an alternative, grant-free non-orthogonal user access schemes have
recently attracted significant attention, where the active users send
their pilot sequences to the BS simultaneously so that the BS can
perform user activity detection, channel estimation, and/or data
decoding in one shot
\cite{zhu,dekorsy1,xu_rao_lau,wunder2,hannak,dekorsy2,wunder_5G,wunder1,zhilin_ICASSP}.
In contrast to the traditional channel estimation process,
the devices in these systems use non-orthogonal pilot sequences,
due to the massive number of devices in the network and limited
coherence time of the wireless channels.
The key observation made in \cite{zhu,dekorsy1,xu_rao_lau,wunder2,hannak,dekorsy2,wunder_5G,wunder1,zhilin_ICASSP}
is that the sparsity in user activity pattern 
allows the formulation of a compressed sensing problem.  For
example, assuming perfect CSI at the BS, a joint user activity and
data detection for code division multiple access (CDMA) systems is
investigated in \cite{zhu,dekorsy1} by exploiting
various sparsity structure. 
When CSI is not available at the BS, user activity detection
and channel estimation is jointly performed in
\cite{xu_rao_lau,wunder2}, where \cite{xu_rao_lau} proposes a modified
Bayesian compressed sensing algorithm in a cloud radio-access network,
and \cite{wunder2} introduces a one-shot random access protocol and
employs a basis pursuit denoising method with a detection error bound
in an orthogonal frequency division multiplexing (OFDM) system.
Moreover, 
\cite{hannak,dekorsy2,wunder_5G,wunder1} perform joint information
decoding in addition based on various compressed sensing techniques.

One of the open issues in the above line of work is the lack of
a rigorous performance analysis for the non-orthogonal multiple-access
scheme for massive connectivity. Toward this end, the recent work of
\cite{zhilin_ICASSP, zhilin_JSAC} proposes the use of the
AMP algorithm for the joint user
activity detection and channel estimation problem, and further shows
that a state evolution analysis \cite{bayati_montanari_amp} of
the AMP algorithm allows an analytic characterization of the missed
detection and false alarm probabilities for device detection. The
analysis of \cite{zhilin_ICASSP,zhilin_JSAC} is however quite involved, especially
when the BS is equipped with multiple antennas. Further,
\cite{zhilin_ICASSP,zhilin_JSAC} analyze device detection performance
only. Its impact on user achievable rate has not been investigated.

This paper makes further progress in this direction by making an
observation that the characterization of device detection performance
and user achievable rates can be simplified substantially in the
massive MIMO regime where the number of BS antennas goes to infinity.
In particular, our asymptotic analysis provides a key insight that
the channel estimation error, rather than detection error, is the
limiting factor in the massive MIMO regime.

We emphasize that while massive MIMO system has been extensively
studied in the literature \cite{marzetta,larsson,debbah,larsson14,debbah12},
conventional analysis focuses on the regime of small number of users
as compared to the number of BS antennas. By contrast, the analysis of
this paper assumes that the number of devices can also go to infinity,
while focusing on the cost of device detection in the overall design.
We also note here that the information theoretic upper bounds for
massive connectivity have been derived in \cite{guo3,yu_ITA} for both
the cases with and without CSI, respectively. However, how to approach
these rate upper bounds in a practical massive connectivity system is
still an open problem.

The main tool in our analysis is the AMP algorithm, which is
originally proposed to solve the single measurement vector (SMV) based
sparse signal recovery problem \cite{donoho_amp}, then generalized to
the multiple measurement vector (MMV) problem in
\cite{baron_MMV,schniter_mmv}, corresponding to the multiple-antenna
case in our setting. The AMP
algorithm is also generalized to the case where the input signal
is not independently and identically distributed (i.i.d.) in \cite{rangan3}. As compared to other compressed sensing techniques, AMP has many appealing advantages.
First, the AMP algorithm has low complexity similar to other iterative thresholding algorithms while achieving the reconstruction power of the
basis pursuit at the same time \cite{donoho_amp}. Second, the performance of the AMP algorithm can be accurately
predicated by state evolution \cite{bayati_montanari_amp}.

\subsection{Main Contributions}

This two-part paper provides an analytical performance characterization of the
two-phase transmission protocol in a single-cell massive connectivity
scenario with massive MIMO, in which the active users send their
non-orthogonal pilot sequences to the BS simultaneously for user
activity detection and channel estimation in the first phase of the
coherence time, then transmit data to the BS for information decoding
in the second phase. By studying certain asymptotic regime where the
number of BS antennas, the number of potential devices, and the number
of active devices all go to infinity, this paper is able to analyze
the performance of user activity detection and channel estimation, and
further to characterize the overall achievable rates while taking the
cost of both user activity detection and channel estimation into
consideration. Specifically, the main contributions of this paper are
as follows.

In Part I of this paper,
we design a minimum mean-squared error (MMSE) denoiser in
the vector AMP algorithm for user activity detection and channel
estimation based on statistical channel information.  By exploiting
the state evolution of vector AMP, we show that the MMSE-based AMP
algorithm is capable of driving the user detection error probability
down to zero as the number of antennas at the BS goes to infinity.
This implies that perfect
user activity detection is possible in a practical IoT/MTC system
if the BS is equipped with a sufficiently large number of antennas.
Further, the statistical distributions of the estimated channels
can also be obtained analytically thanks to the state evolution.
These results are used in Part II of this paper \cite{LiuPart2} to characterize
achievable transmission rates of massive connectivity.

\subsection{Organization}
The rest of Part I of this paper is organized as follows.
Section \ref{sec:System Model} describes the system model for massive
connectivity and introduces the two-phase transmission protocol for
user detection, channel estimation, and data transmission.
Section \ref{sec:Phase I} presents the vector AMP based user activity
detection and channel estimation scheme;
Section \ref{sec:Analysis of AMP Algorithm} analyzes its
detection error probability and channel estimation error performance;
Section \ref{sec:AMP in Massive MIMO Regime} further provides an
asymptotic analysis in the massive MIMO regime.
Section \ref{sec:Numerical Examples} provides the numerical
simulation results pertaining to device detection and channel
estimation error. Finally, Section \ref{sec:Conclusion} concludes
this Part I of the paper.


\subsection{Notation}

Scalars are denoted by lower-case letters, vectors by bold-face
lower-case letters, and matrices by bold-face upper-case letters.
The identity matrix and the all-zero matrix of appropriate
dimensions are denoted as $\mv{I}$ and $\mv{0}$, respectively.
For a matrix $\mv{M}$ of arbitrary size, $\mv{M}^{H}$ and $\mv{M}^{T}$
denote its conjugate transpose and transpose, respectively.
The expectation operator is denoted as $\mathbb{E}[\cdot]$.
The distribution of a circularly symmetric complex Gaussian (CSCG)
random vector with mean $\mv{x}$ and covariance matrix $\mv{\Sigma}$
is denoted by $\mathcal{CN}(\mv{x},\mv{\Sigma})$; 
the space of complex matrices of size $m\times n$ is denoted as
$\mathbb{C}^{m \times n}$.

\section{System Model}\label{sec:System Model}

Consider the uplink of a single-cell cellular network consisting
of $N$ users, denoted by the set $\mathcal{N}=\{1,\cdots,N\}$.
It is assumed that the BS is equipped with $M$ antennas, while
each user is equipped with one antenna. The complex
uplink channel vector from user $n$ to the BS is denoted by
$\mv{h}_n\in \mathbb{C}^{M\times 1}$, $n=1,\cdots,N$. This paper
adopts a block-fading model, in which all the channels follow
independent quasi-static flat-fading within a block of coherence
time, where $\mv{h}_n$'s remain constant, but vary independently
from block to block. Moreover, we model the channel vector
$\mv{h}_n$ as $\mv{h}_n=\sqrt{\beta_n}\mv{g}_n$, $\forall n$, where
$\mv{g}_n\sim \mathcal{CN}(\mv{0},\mv{I})$ denotes the Rayleigh
fading component, and $\beta_n$ denotes the path-loss and shadowing
component.  Therefore, we have
$\mv{h}_n\sim \mathcal{CN}(\mv{0},\beta_n\mv{I}_n)$, $\forall n$.
The path-loss and shadowing components depend on the user location
and are assumed to be known at the BS.\footnote{In this paper we mainly focus on the scenario where the devices are stationary, e.g., home security systems, smart metering systems, etc.}

The sporadic nature of user traffic is modeled as follows. We assume
that the users are synchronized and each user decides in each
coherence block whether or not to access the channel with probability
$\epsilon$ in an i.i.d.\ manner.
Thus, within each coherence block only a subset of the users are active.
In each particular block, we define the user activity indicator for
user $n$ as follows:
\begin{align}\label{eqn:user activity indicator}
\alpha_n=\left\{\begin{array}{ll}1, & {\rm if ~ user} ~ n ~ {\rm is ~ active}, \\ 0, & {\rm otherwise},\end{array} \right. ~~~ \forall n,
\end{align}
so that
${\rm Pr}(\alpha_n=1)=\epsilon$, ${\rm Pr}(\alpha_n=0)=1-\epsilon$, $\forall n$.
Further, we define the set of active users within a coherence
block as
\begin{align}\label{sec:active users}
\mathcal{K}=\{n:\alpha_n=1, n=1,\cdots,N\}.
\end{align}
We denote the number of active users as $K$, i.e., $K=|\mathcal{K}|$.
The overall channel input-output relationship is modeled as:
\begin{align}\label{eqn:channel}
\mv{y}= \sum_n \mv{h}_n \alpha_n x_n + \mv{z} =
\sum\limits_{k\in \mathcal{K}}\mv{h}_k x_k+\mv{z},
\end{align}
where $x_n \in \mathbb{C}$, $\mv{y} \in \mathbb{C}^{M\times 1}$, and $\mv{z}
\in \mathbb{C}^{M\times 1}$ respectively are the user transmitted signal,
the channel output at the BS, and the additive white
Gaussian noise (AWGN) vector distributed as
$\mathcal{CN}(\mv{0}, \sigma^2 \mv{I})$.
For simplicity, this paper assumes no power adaptation so that all
the devices transmit at a constant power $\rho$, i.e.,
\begin{equation}
\mathbb{E}|x_n|^2 = \rho.
\end{equation}
Further, we do not account for intercell interference.  The objective
at the BS is to detect which users are active and to decode their
transmitted messages within each coherence block.

This paper adopts a grant-free multiple-access scheme as shown in
Fig.~\ref{fig2}, in which each coherence block of length $T$ symbols is
divided into two phases. In the first phase, the active users send
their pilot sequences of length $L$ symbols to the BS synchronously,
and the BS jointly detects the user activities, i.e., $\alpha_n$'s, as
well as the active users' channels, i.e., $\mv{h}_n$'s,
$\forall n \in \mathcal{K}$. In the second phase, the active users
send their data to the BS using the remaining $T-L$ symbols, and the
BS decodes these messages based on the knowledge of user activities
and channels obtained in the first phase.

For the massive connectivity scenario with a large number of potential
devices, the length of pilot sequence is typically smaller than the
total number of devices, i.e., $L<N$. In this case, it is not possible
to assign mutually orthogonal sequences to all the users. Following the pioneering works on AMP \cite{donoho_amp,baron_MMV,schniter_mmv,rangan3,bayati_montanari_amp}, this paper
assumes that pilot sequences are generated randomly, i.e., each user
$n$ is assigned a unique pilot sequence
\begin{equation}
\mv{a}_n=[a_{n,1},\cdots,a_{n,L}]^T\in \mathbb{C}^{L\times 1},
\end{equation}
whose entries are generated from i.i.d.\ complex Gaussian distribution with
zero mean and variance $1/L$, i.e.,
\begin{equation}
a_{n,l}\sim \mathcal{CN}\left(0,\frac{1}{L}\right),
\end{equation}
so that each pilot sequence has unit norm, i.e., $\|\mv{a}_n\|^2=1$,
as $L\rightarrow \infty$. It is further assumed that the pilot
sequences of all the users are known by the BS.

To facilitate analysis, this paper considers a certain asymptotic
regime where $N \rightarrow \infty$, so that $K \rightarrow
\epsilon N$, and the empirical distribution of $\beta_1,\cdots,\beta_N$
converges to a fixed distribution denoted by $p_\beta$.





\section{User Activity Detection and Channel Estimation via AMP}
\label{sec:Phase I}
In this section, we show that the AMP algorithm can be used in the
first phase for user activity detection and channel estimation by
exploiting the sparsity in user activity.

\subsection{Device Activity Detection and Channel Estimation as a
Compressed Sensing Problem}

Consider the first phase of massive device transmission in which each
user sends its pilot sequence synchronously through the channel. Define
$\rho^{\rm pilot}$ as the identical transmit power of the active users
in the first transmission phase. The transmit signal of user $n$ can be
expressed as $\alpha_n \sqrt{\xi } \mv{a}_n$, where $\xi  = L \rho^{\rm pilot} $
denotes the total transmit energy of each active user in the first phase.\footnote{In practice, the pilot sequence length $L$ is large but finite. It is thus possible that the power of the randomly generated pilot $\sqrt{\xi }\mv{a}_n$ is slightly larger than $\xi$ for some users. To satisfy the transmit power constraint, in practice we can generate the pilot sequences based on a modified distribution: $a_{n,l}\sim \mathcal{CN}(0,1/L-\zeta)$. With a careful choice of $\zeta$, the probability that one pilot sequence violates the power constraint can be close to zero.} The received signal at the BS is then
\begin{align}\label{eqn:received signal phase 1}
\mv{Y} =\sqrt{\xi }\sum\limits_{n\in \mathcal{N}}\alpha_n\mv{a}_n\mv{h}_n^T+\mv{Z},
\end{align}
where $\mv{Y} \in \mathbb{C}^{L\times M}$ is the matrix of
received signals across $M$ antennas over $L$ symbols, and
$\mv{Z}=[\mv{z}_1,\cdots,\mv{z}_M]$ with
$\mv{z}_m \sim \mathcal{CN}(\mv{0},\sigma^2\mv{I})$, $\forall m$,
is the independent AWGN at the BS.
Now define
\begin{equation}
\mv{A}=[\mv{a}_1, \cdots, \mv{a}_N].
\end{equation}
Let $\mv{x}_n=\alpha_n\mv{h}_n$ and define
\begin{equation}
\mv{X}=[\mv{x}_1, \cdots, \mv{x}_N]^T.
\end{equation}
Then, the training phase can be modeled as the following matrix
equation
\begin{equation}
\label{eqn:channel_training}
\mv{Y} =\sqrt{\xi }\mv{A}\mv{X}+\mv{Z},
\end{equation}
where the rows of the matrix $\mv{X}$ follow 
a Bernoulli Gaussian distribution:
\begin{align}\label{eqn:distribution}
p_{\mv{x}_n}=(1-\epsilon)\delta_0+\epsilon p_{\mv{h}_n}, ~~~ \forall n.
\end{align}
Here, $\delta_0$ denotes the point mass measure at zero, and
$p_{\mv{h}_n}$ denotes the distribution of user $n$'s channel
$\mv{h}_n\sim \mathcal{CN}(\mv{0},\beta_n\mv{I})$.

The goal for the BS in the first phase is to detect the user activities and to
estimate the user channels by recovering $\mv{X}$ based on the noisy
observation $\mv{Y}$. 
As $\mv{X}$ is row sparse, i.e., many $\mv{x}_n$'s are zero, such a
reconstruction problem is a compressed sensing problem. Further, as
the sparsity pattern is sensed at multiple antennas, this is an MMV
compressed sensing setup.

Among many powerful compressed sensing techniques, this paper adopts a
low-complexity AMP algorithm to recover the row-sparse matrix $\mv{X}$.
Before proceeding to evaluate its performance
for user activity detection and channel estimation, we first briefly review the vector
version of the AMP algorithm in next subsection.

\subsection{Vector AMP Algorithm}


The vector AMP algorithm is first proposed in \cite{baron_MMV}.
This paper considers a version of the algorithm proposed in
\cite{zhilin_JSAC} that aims to provide an estimate
$\mv{\hat{X}}(\mv{Y})$ based on $\mv{Y}$ that
minimizes the mean-squared error (MSE)
\begin{equation}
\label{eqn:MSE}
{\rm MSE} =
\mathbb{E}_{\mv{X} \mv{Y}} || \mv{\hat{X}}(\mv{Y}) - \mv{X} ||^2_2
\end{equation}
assuming the prior distribution (\ref{eqn:distribution}) and the
channel model (\ref{eqn:channel_training}).
The algorithm is based on an approximation of the message passing
algorithm for solving the above problem.

Starting with $\mv{X}^0=\mv{0}$ and $\mv{R}^0=\mv{Y}$,
the general form of the vector AMP algorithm proceeds at each iteration
as \cite{rangan3,baron_MMV,zhilin_JSAC}:
\begin{align}
\mv{x}_n^{t+1} & =\eta_{t,n}((\mv{R}^t)^H\mv{a}_n+\mv{x}_n^t), \label{eqn:AMP 1} \\
\mv{R}^{t+1} & =\mv{Y}-\mv{A}\mv{X}^{t+1}+\frac{N}{L}\mv{R}^t\sum\limits_{n=1}^N\frac{\eta_{t,n}'((\mv{R}^t)^H\mv{a}_n+\mv{x}_n^t)}{N}, \label{eqn:AMP 2}
\end{align}where $t=0,1,\cdots$ is the index of the iteration, $\mv{X}^t=[\mv{x}_1^t,\cdots,\mv{x}_N^t]^T$ is
the estimate of $\mv{X}$ at iteration $t$, and $\mv{R}^t=[\mv{r}_1^t,\cdots,\mv{r}_L^t]^T\in \mathbb{C}^{L\times M}$
denotes the corresponding residual.
Intuitively, the algorithm performs in (\ref{eqn:AMP 1}) a matching
filtering of the residual for each user $n$ using its
pilot sequence, followed by a denoising step using an
appropriately designed denoiser $\eta_{t,n}(\cdot):
\mathbb{C}^{M\times 1}\rightarrow \mathbb{C}^{M\times 1}$.
The residual is then updated in (\ref{eqn:AMP 2}), but corrected
with a so-called Onsager term involving $\eta_{t,n}'(\cdot)$,
the first-order derivative of $\eta_{t,n}(\cdot)$.


A remarkable property of the AMP algorithm is that when applied to the
compressed sensing problem with the entries of the sensing matrix
$\mv{A}$ generated from i.i.d.\ Gaussian distribution, its detection
performance in certain asymptotic regime can be accurately predicted by the
so-called \emph{state evolution}. The asymptotic regime is when
$L,K,N\rightarrow \infty$, while their ratios converge to some fixed
positive values $N/L \rightarrow \omega$ and $K/N \rightarrow
\epsilon$ with $\omega, \epsilon \in (0,\infty)$, while keeping
the total transmit power fixed at $\xi $. Note that we fix
the total transmit power rather than allowing it to scale with $L$
here in this hypothetical asymptotic system in order to carry out the
state evolution analysis. (This implies that the per-symbol power goes
down to zero in this hypothetical asymptotic regime.) The analysis is
then used to predict the system performance at finite (but large) $L, K, N$ and
$\xi =L\rho^{\rm pilot}$. As shown in \cite{zhilin_ICASSP}, this approach is
found to corroborate very well with simulation results.

Specifically, given $\beta \sim p_\beta$, define a random vector
$\mv{X}_\beta\in \mathbb{C}^{M\times 1}$ with a
distribution $(1-\epsilon)\delta_0+\epsilon p_{\mv{h}_\beta}$,
where $p_{\mv{h}_\beta}$ denotes the distribution
$\mv{h}_\beta\sim \mathcal{CN}(\mv{0},\beta\mv{I})$.
Let $\mv{V}\in \mathbb{C}^{M\times 1}\sim \mathcal{CN}(\mv{0},\mv{I})$
be independent of $\mv{X}_\beta$. Then, the state evolution is
the following recursion for $t\geq 0$
\cite{bayati_montanari_amp,baron_MMV,rangan3, zhilin_JSAC}:
\begin{align}\label{eqn:state evolution}
\mv{\Sigma}_{t+1}=\frac{\sigma^2}{\xi }\mv{I}+
	\omega \mathbb{E}\bigg[&(\eta_{t,\beta}(\mv{X}_\beta+\mv{\Sigma}_t^{\frac{1}{2}}\mv{V})-\mv{X}_\beta)\nonumber \\ & (\eta_{t,\beta}(\mv{X}_\beta+\mv{\Sigma}_t^{\frac{1}{2}}\mv{V})-\mv{X}_\beta)^H\bigg],
\end{align}
where $\mv{\Sigma}_t$ is referred to as the state, and
the expectation is over $\beta$, $\mv{X}_\beta$ and $\mv{V}$.
Note that in the above equation,
$\eta_{t,n}(\cdot)$ is replaced by $\eta_{t,\beta}(\cdot)$
for convenience, since $n$ and $\beta_n$ are interchangeable.
Moreover, the initial point to the above state evolution is
the noise covariance matrix after the first matched filtering, i.e.,
\begin{align}\label{eqn:state evolution initial point}
\mv{\Sigma}_0=\frac{\sigma^2}{\xi }\mv{I}+\omega\mathbb{E}[\mv{X}_\beta\mv{X}_\beta^H].
\end{align}

Define a set of random vectors $\hat{\mv{X}}_{t,n}=\mv{X}_n+\mv{\Sigma}_t^{\frac{1}{2}}\mv{V}_n$, $\forall n$,
where $\mv{X}_n\in \mathbb{C}^{M\times 1}$ follows the distribution given in (\ref{eqn:distribution}),
and $\mv{V}_n\in \mathbb{C}^{M\times 1}
\sim \mathcal{CN}(\mv{0},\mv{I})$ is independent of $\mv{X}_n$.
The state evolution analysis says that in the vector AMP algorithm, applying the denoiser
to $(\mv{a}_n^H\mv{R}^t)^H+\mv{x}_n^t$ as shown in (\ref{eqn:AMP 1}) is statistically equivalent to applying the denoiser to
\begin{align}\label{eqn:equivalent channel}
\hat{\mv{x}}_{t,n}=\mv{x}_n+\mv{\Sigma}_t^{\frac{1}{2}}\mv{v}_n=\alpha_n\mv{h}_n+\mv{\Sigma}_t^{\frac{1}{2}}\mv{v}_n, ~~~ \forall n,
\end{align}
where the distributions of $\hat{\mv{x}}_{t,n}$ and $\mv{v}_n$ are
captured by the random vectors $\hat{\mv{X}}_{t,n}$ and $\mv{V}_n$ \cite{bayati_montanari_amp,baron_MMV,rangan3}.

\subsection{MMSE Denoiser Design for Vector AMP}

The key advantage of the equivalent signal model given in
(\ref{eqn:equivalent channel}) is the decoupling of the estimation
between different users, which allows us to design the denoiser
$\eta_{t,n}(\cdot)$ in the vector AMP algorithm to minimize the
MSE (\ref{eqn:MSE}) based on the above decoupled
signal model. Specifically,
in the $t$th iteration of the AMP algorithm, the MMSE denoiser
$\eta_{t,n}(\cdot)$ is given by the conditional expectation
$\mathbb{E}[\mv{X}_n|\hat{\mv{X}}_{t,n}]$. This denoiser has been
derived in \cite{baron_MMV, zhilin_JSAC}, but is re-derived in Appendix
\ref{app:lemmas} and expressed below in a form that highlights its
structural dependence in $M$:
\begin{align}\label{eqn:MMSE denoiser}
\eta_{t,n}(\hat{\mv{x}}_{t,n})& =
\mathbb{E}[\mv{X}_n|\hat{\mv{X}}_{t,n}=\hat{\mv{x}}_{t,n}] \nonumber \\ & =\phi_{t,n}\beta_n(\beta_n\mv{I}+\mv{\Sigma}_t)^{-1}\hat{\mv{x}}_{t,n}, ~~~ \forall t,n,
\end{align}where
\begin{align}
& \phi_{t,n}=\frac{1}{1+\frac{1-\epsilon}{\epsilon}{\rm exp}\left(-M\left(\pi_{t,n}-\psi_{t,n}\right)\right)}, \label{eqn:threshold}  \\
& \pi_{t,n}=\frac{\hat{\mv{x}}_{t,n}^H(\mv{\Sigma}_t^{-1}-(\mv{\Sigma}_t+\beta_n\mv{I})^{-1})\hat{\mv{x}}_{t,n}}{M}, \label{eqn:tau} \\
& \psi_{t,n}=\frac{\log \det(\mv{I}+\beta_n\mv{\Sigma}_t^{-1})}{M}. \label{eqn:psi}
\end{align}

Examining the functional form of the MMSE denoiser
(\ref{eqn:MMSE denoiser})-(\ref{eqn:psi}), it is worthwhile to note
that if all the users are active, i.e., $\epsilon=1$, it follows
that $\phi_{t,n}=1$, $\forall n$, in which case the MMSE denoiser
given in (\ref{eqn:MMSE denoiser}) reduces to the linear MMSE
estimator:
$\eta_{t,n}(\hat{\mv{x}}_{t,n})=\beta_n(\beta_n\mv{I}+\mv{\Sigma}_t)^{-1}\hat{\mv{x}}_{t,n}$,
which is widely used for channel training when the user activity is known \cite{marzetta,larsson,debbah}.
When the device activity needs to be detected, the above MMSE denoiser
is a non-linear function of $\hat{\mv{x}}_{t,n}$ due to the functional form of $\phi_{t,n}$.

\subsection{State Evolution for MMSE Denoiser Based Vector AMP}

The general state evolution of the AMP algorithm as in
(\ref{eqn:state evolution}) applies to any arbitrary denoiser
$\eta_{t,n}(\cdot)$. With the MMSE denoiser (\ref{eqn:MMSE denoiser}),
the state evolution can be considerably simplified.

\begin{theorem}\label{theorem7}
Consider the MMSE denoiser based vector AMP algorithm for device
detection and channel estimation in the asymptotic regime in which
the number of users $N$, the number of active users $K$, and the
length of the pilot sequences $L$ all go to infinity, while their
ratios converge to some fixed positive values, i.e., $N/L \rightarrow \omega$
and $K/N \rightarrow \epsilon$ with $\omega, \epsilon \in (0,\infty)$,
and while keeping the total transmit power $\xi $ fixed.
The matrix $\mv{\Sigma}_t$ in the state evolution (\ref{eqn:state evolution})
always stays as a diagonal matrix with identical diagonal entries
after each iteration, i.e.,
\begin{align}\label{eqn:Sigma diagonal}
\mv{\Sigma}_t=\tau_t^2\mv{I}, ~~~ \forall t\geq 0.
\end{align}
In this case, the signal model given in (\ref{eqn:equivalent channel})
reduces to
\begin{align}\label{eqn:equivalent channel 1}
\hat{\mv{x}}_{t,n}=\mv{x}_n+\tau_t\mv{v}_n.
\end{align}
Moreover, the MMSE
denoiser given in (\ref{eqn:MMSE denoiser}) reduces to
\begin{align}\label{eqn:MMSE denoiser 1}
\eta_{t,n}(\hat{\mv{x}}_{t,n})= \phi_{t,n}
\left(\frac{\beta_n}{\beta_n+\tau_t^2}\right)\hat{\mv{x}}_{t,n}, ~~~ \forall t,n,
\end{align}
where $\phi_{t,n}$ is given in (\ref{eqn:threshold}), while
$\pi_{t,n}$ and $\psi_{t,n}$ are respectively given by
\begin{align}
&
\pi_{t,n}=
\left(\frac{1}{\tau_t^2}-\frac{1}{\tau_t^2+\beta_n}\right)
\frac{ \hat{\mv{x}}_{t,n}^H\hat{\mv{x}}_{t,n}}{M}, \label{eqn:pi 2} \\
& \psi_{t,n}=\log\left(1+\frac{\beta_n}{\tau_t^2}\right). \label{eqn:psi 2}
\end{align}
At last, $\tau_t^2$ can be iteratively obtained as follows
\begin{align}
& \tau_0^2=\frac{\sigma^2}{\xi}+\omega \epsilon \mathbb{E}_\beta[\beta],
	\label{eqn:state evolution0}\\
& \tau_{t+1}^2=\frac{\sigma^2}{\xi}+\omega\epsilon\mathbb{E}_\beta\left[\frac{\beta\tau_t^2}{\beta+\tau_t^2}\right]+\omega\mathbb{E}_\beta[\vartheta_{t,\beta}(\tau_t^2)], ~ t\geq 0, \label{eqn:state evolution fixed M}
\end{align}
with the term $\vartheta_{t,\beta}(\tau_t^2)$ expressed as
\begin{align}\label{eqn:vartheta 1}
\vartheta_{t,\beta}(\tau_t^2)=\frac{1}{M}\mathbb{E}_{\hat{\mv{X}}_{t,\beta}}
\left[\phi_{t,\beta}(1-\phi_{t,\beta})\frac{\beta^2}{(\beta+\tau_t^2)^2}\hat{\mv{X}}_{t,\beta}^H\hat{\mv{X}}_{t,\beta}\right].
\end{align}
Here, $\hat{\mv{X}}_{t,\beta}$ is the random vector that captures
the distribution of the signal $\hat{\mv{x}}_{t,n}$ given in
(\ref{eqn:equivalent channel 1}), and $\phi_{t,\beta}$ captures the
distribution of $\phi_{t,n}$, which is implicitly a function of
$\hat{\mv{x}}_{t,n}$.
\end{theorem}

\begin{IEEEproof}
Please refer to Appendix \ref{appendix11}.
\end{IEEEproof}

The key observation of Theorem \ref{theorem7} is that because each
user's channels across the multiple receive antennas at the BS are
assumed to be uncorrelated, the residual noise in (\ref{eqn:equivalent
channel}) remains uncorrelated across the antennas. This is in spite of
the fact that the vector AMP algorithm involves non-linear processing
in $\phi_{t,n}$ as in (\ref{eqn:MMSE denoiser}).
This scalar form of the state evolution significantly simplifies
performance analysis of the device activity detection and channel
estimation.


\subsection{Device Detection and Channel Estimation by Vector AMP}


We are now ready to state the proposed device activity detector and
channel estimator. Observe that the functional form of $\phi_{t,n}$
as given in (\ref{eqn:MMSE denoiser}) is such that for large $M$, we
have that $\phi_{t,n}$ tends to $1$ if $\pi_{t,n} > \psi_{t,n}$ and $0$
if $\pi_{t,n} < \psi_{t,n}$. As a result, the AMP algorithm suggests
that it is reasonable to adopt a threshold strategy for user activity
detection, i.e., to declare a user as active or not simply
based on whether $\pi_{t,n}$ exceeds the threshold of $\psi_{t,n}$.
Using (\ref{eqn:pi 2})-(\ref{eqn:psi 2}) and the scalar form of the
AMP state evolution, the proposed device activity detector and channel
estimator are as follows.

\begin{figure*}[hb]
\hrulefill
\setcounter{equation}{39}
\begin{align}
& \upsilon_{t,k}(M)=\frac{1}{M}\mathbb{E}\left[ \phi_{t,k}^2
\left(\frac{\beta_k}{\beta_k+\tau_t^2}\right)^2
(\mv{h}_k+\tau_t\mv{v}_k)^H(\mv{h}_k+\tau_t\mv{v}_k)\right], \label{eqn:channel 3} \\
& \Delta
\upsilon_{t,k}(M)=\frac{1}{M}\mathbb{E}\left[\left(\phi_{t,k}\frac{\beta_k}{\beta_k+\tau_t^2}(\mv{h}_k+\tau_t\mv{v}_k)-\mv{h}_k\right)^H
\left(\phi_{t,k}\frac{\beta_k}{\beta_k+\tau_t^2}(\mv{h}_k+\tau_t\mv{v}_k)-\mv{h}_k\right) \right]. \label{eqn:channel 4}
\end{align}
\setcounter{equation}{29}
\end{figure*}

\begin{definition}
The MMSE vector AMP algorithm based device activity detector is
defined as the following threshold-based detector.
After $t$ iterations of AMP, compute $\tau_t^2$ according to
(\ref{eqn:state evolution0})-(\ref{eqn:state evolution fixed M}).
For each user $n$, compare $\pi_{t,n}$ and $\psi_{t,n}$ as defined
in (\ref{eqn:pi 2})-(\ref{eqn:psi 2}), i.e.,
\begin{align}
& \varpi(\mv{x}_n^t,\mv{R}^t) = \nonumber \\ & \left\{\begin{array}{ll}1, & {\rm if} ~
((\mv{R}^t)^H\mv{a}_n+\mv{x}_n^t)^H ((\mv{R}^t)^H\mv{a}_n+\mv{x}_n^t) > \theta_{t,n}, \\
0, & {\rm if} ~
((\mv{R}^t)^H\mv{a}_n+\mv{x}_n^t)^H ((\mv{R}^t)^H\mv{a}_n+\mv{x}_n^t) < \theta_{t,n},
\end{array}\right. \label{eqn:detection function}
\end{align}
with a threshold $\theta_{t,n} = 
M \log\left(1+\frac{\beta_n}{\tau_t^2}\right) \left/
\left(\frac{1}{\tau_t^2}-\frac{1}{\tau_t^2+\beta_n}\right).\right.
$
Further, given that a device $k$ is declared active, its channel is
estimated as:
\begin{align}\label{eqn:estimator function}
\mv{\hat{h}}_{t,k} = \mv{x}_k^t.
\end{align}
\end{definition}

We remark that the MMSE denoiser given in (\ref{eqn:MMSE denoiser 1}) is a scaled version of the observation. As a result, the complexity of the vector AMP algorithm introduced in (\ref{eqn:AMP 1}) and (\ref{eqn:AMP 2}) mainly comes from the matrix multiplication $\mv{A}\mv{X}^{t+1}$. Since $\mv{A}\in \mathbb{C}^{L\times N}$ and $\mv{X}^{t+1}\in \mathbb{C}^{N\times M}$, the complexity of the AMP algorithm is in the order of $\mathcal{O}(LNM)$ per iteration.

\section{Performance of Vector AMP Based Detector}
\label{sec:Analysis of AMP Algorithm}


In this section, we
analyze the performance of the above MMSE vector AMP based
device activity detector in terms of missed detection and false alarm
error probabilities, as well as the associated channel estimation
error based on the state evolution, as functions of the number of
receive antennas $M$ at the BS.  The results of this section pertain
to finite $M$.  Asymptotic result with $M$ going to infinity is
presented in the next section.


\subsection{Probabilities of Missed Detection and False Alarm}


We are ready to
examine the error probability for the MMSE vector AMP
based detector as defined in (\ref{eqn:detection function}).
Let the missed detection and false alarm probabilities of user $n$
after the $t$th iteration of the MMSE denoiser based AMP algorithm
be defined as
\begin{equation}\label{eqn:missed detection probability}
P^{\rm MD}_{t,n}(M)={\rm Pr}( \varpi(\mv{x}_n^t,\mv{R}^t)=0 | \alpha_n=1),
\end{equation}
and
\begin{equation}\label{eqn:false alarm probability}
P^{\rm FA}_{t,n}(M)={\rm Pr}( \varpi(\mv{x}_n^t,\mv{R}^t)=1 | \alpha_n=0),
\end{equation}
respectively, as functions of $M$, the number of antennas at the BS.
The following theorem characterizes $P^{\rm MD}_{t,n}(M)$ and
$P^{\rm FA}_{t,n}(M)$ analytically in terms of $\tau_t^2$ and $M$.

\begin{theorem}\label{theorem15}
Consider the device activity detector for massive connectivity
(\ref{eqn:detection function}), based on
the vector AMP algorithm with MMSE denoiser. Fix the number of BS
receive antennas $M$, and consider the asymptotic regime in which
the number of users $N$, the number of active users $K$, and the
length of the pilot sequences $L$ all go to infinity, while their
ratios converge to some fixed positive values, i.e., $N/L \rightarrow \omega$
and $K/N \rightarrow \epsilon$ with $\omega, \epsilon \in (0,\infty)$,
and while keeping the total transmit power $\xi $ fixed. 
After $t$ iterations of the AMP algorithm, the probabilities of missed
detection and false alarm of user $n$ can be expressed, respectively, as
\begin{align}\label{eqn:missed detection probability 1}
P^{\rm MD}_{t,n}(M)=\frac{\underline{\Gamma}\left(M,b_{t,n}M\right)}{\Gamma(M)},
\end{align}and
\begin{align}\label{eqn:false alarm probability 1}
P^{\rm FA}_{t,n}(M)=\frac{\bar{\Gamma}\left(M,c_{t,n}M\right)}{\Gamma(M)},
\end{align}where $\Gamma(\cdot)$, $\underline{\Gamma}(\cdot,\cdot)$, and $\bar{\Gamma}(\cdot,\cdot)$ denote the Gamma function, lower
incomplete Gamma function, and upper incomplete Gamma function \cite{Gautschi}, respectively, and
\begin{align}
& b_{t,n}= 
\frac{\tau_t^2}{\beta_n} \log\left(1+\frac{\beta_n}{\tau_t^2}\right), \label{eqn:b1}\\
& c_{t,n}= 
\left(\frac{\beta_n + \tau_t^2}{\beta_n} \right)\log\left(1+\frac{\beta_n}{\tau_t^2}\right),
\label{eqn:b2}
\end{align}
and $\tau_t^2$ is given by (\ref{eqn:state evolution0})-(\ref{eqn:state evolution fixed M}).
\end{theorem}
\begin{IEEEproof}
Please refer to Appendix \ref{appendix15}.
\end{IEEEproof}

The above analysis of detection probabilities of error hinges upon the
Gaussian signal model of the AMP state evolution
(\ref{eqn:equivalent channel}). As consequence of the scalar signal
model (\ref{eqn:equivalent channel 1}), the proposed device
activity detector (\ref{eqn:detection function}) becomes
a threshold detector on a $\chi^2$-distribution as defined in
(\ref{eqn:pi 2}). This allows both missed detection and false alarm
probabilities to be characterized analytically.

An important observation is that due to the fact that
$a > \log(1+a) > \frac{a}{1+a}$ for $a>0$, we have that
$b_{t,n} < 1$ and $c_{t,n}> 1$. As result, one can show that both
error probabilities eventually go to zero as $M \rightarrow \infty$.
This asymptotic behavior is explored in more detail in the next section.

Finally, we note that
at the convergence of the AMP algorithm, $\tau_t^2$
converges to the fixed-point solution to (\ref{eqn:state evolution
fixed M}), i.e., $\tau_\infty^2$. The detection error
probabilities may then be expressed as (\ref{eqn:missed detection
probability 1})-(\ref{eqn:false alarm probability 1}) with $\tau_t^2$
replaced by $\tau_\infty^2$.

\subsection{Analysis of Channel Estimation Error}
The state evolution analysis of AMP further allows us to evaluate the
channel estimation performance.
For any active user $k\in \mathcal{K}$, the estimated channel
$\hat{\mv{h}}_{t,k}$ is as defined in (\ref{eqn:estimator function}),
with $\Delta \mv{h}_{t,k}=\mv{h}_{t,k}-\hat{\mv{h}}_{t,k}$ denoting
the corresponding channel estimation error. The following
theorem characterizes the covariance matrices of the estimated
channels and the channel estimation errors of the active
users as function of the number of BS antennas $M$.

\begin{theorem}\label{theorem9}
Consider the channel estimator (\ref{eqn:estimator function}) for
massive connectivity, based on the vector AMP algorithm with MMSE
denoiser. Fix the number of BS receive antennas $M$, and consider the
asymptotic regime in which the number of users $N$, the number of
active users $K$, and the length of the pilot sequences $L$ all go to
infinity, while their ratios converge to some fixed positive values,
i.e., $N/L \rightarrow \omega$ and $K/N \rightarrow \epsilon$ with
$\omega, \epsilon \in (0,\infty)$, and while keeping the total transmit
power $\xi $ fixed. For each active user $k\in \mathcal{K}$, the
covariance matrices of its estimated channel and channel estimation
error are given, respectively, by
\begin{align}
& {\rm Cov}(\hat{\mv{h}}_{t,k},\hat{\mv{h}}_{t,k})
	= \upsilon_{t,k}(M) \mv{I}, \label{eqn:channel 1} \\
& {\rm Cov}(\Delta \mv{h}_{t,k}, \Delta \mv{h}_{t,k})
	= \Delta \upsilon_{t,k} (M)\mv{I}, \label{eqn:channel 2}
\end{align}
where $\upsilon_{t,k}(M)$ and $\Delta \upsilon_{t,k}(M)$ are
given in (\ref{eqn:channel 3}) and (\ref{eqn:channel 4})
at the bottom of the page, where the expectation is over both the channel
$\mv{h}_k$ and the residual noise $\mv{v}_k$ modeled as Gaussian
random variables, and $\tau_t^2$ is given by state evolution
(\ref{eqn:state evolution0})-(\ref{eqn:state evolution fixed M}).
\end{theorem}

\setcounter{equation}{41}

\begin{IEEEproof}
Please refer to Appendix \ref{appendix9}.
\end{IEEEproof}

As already noted earlier, at the convergence of the AMP algorithm,
$\tau_t^2$ converges to $\tau_\infty^2$. Then, the channel estimation
error converges to (\ref{eqn:channel 1})-(\ref{eqn:channel 2}) with
$\tau_t^2$ replaced by $\tau_\infty^2$.

\section{Asymptotic Performance with Massive MIMO}
\label{sec:AMP in Massive MIMO Regime}

We have so far characterized the probabilities of missed detection and
false alarm as well as the covariance matrices of the estimated
channels and channel estimation errors of the proposed MMSE denoiser
based AMP detector in the asymptotic regime where $N, K, L$ all go to
infinity with fixed ratios, but with fixed $M$. In this section,
we further let $M$ go to infinity and study the asymptotic massive MIMO
regime to gain further insight.

\subsection{Asymptotic Performance of User Activity Detection}

A main result of this paper is to show that the proposed device
activity detector performs well in the massive MIMO regime.
The intuitive reason behind this result is that the threshold
detection (\ref{eqn:detection function}) involves a comparison
between $\pi_{t,n}$ as in (\ref{eqn:pi 2}) with $\psi_{t,n}$
as in (\ref{eqn:psi 2}).  According to the AMP state evolution
signal model (\ref{eqn:equivalent channel 1}), as $M \rightarrow \infty$,
by the law of large numbers, we have
\begin{eqnarray}
\pi_{t,n} & \rightarrow &
\left\{ \begin{array}{ll}
\left(\frac{1}{\tau_t^2}-\frac{1}{\tau_t^2+\beta_n}\right) (\tau_t^2+ \beta_n)
& {\rm if\ } \alpha_n = 1 \\
\left(\frac{1}{\tau_t^2}-\frac{1}{\tau_t^2+\beta_n}\right) \tau_t^2
& {\rm if\ } \alpha_n = 0
\end{array}\right. \nonumber
\end{eqnarray}
which simplifies to
$\pi_{t,n} \rightarrow \beta_n/\tau_t^2$ if user $n$ is active, and
$\pi_{t,n} \rightarrow \beta_n/(\tau_t^2+\beta_n)$ if it is not.
Now compare with
$\psi_{t,n} = \log(1+\beta_n/\tau_t^2)$. Using the fact that
$a > \log(1+a) > \frac{a}{1+a}$ for all $a>0$, we always have
\begin{equation}
\frac{\beta_n}{\tau_t^2} > \log\left(1+\frac{\beta_n}{\tau_t^2}\right)
> \frac{\beta_n}{\tau_t^2 + \beta_n}
\end{equation}
as long as $\beta_n > 0$ and $\tau_t^2 < \infty$. Thus, we have
that asymptotically as $M \rightarrow \infty$, it is always true that
$\pi_{t,n} >\psi_{t,n}$ when $\alpha_n=1$ and $\pi_{t,n}<\psi_{t,n}$
when $\alpha_n=0$. In other words, the proposed detector always makes
the correct decision in the massive MIMO regime.

The above argument can be made more precise by utilizing the analytic
probabilities of missed detection and false alarm expressions as
characterized in Theorem \ref{theorem15} to show that the error
probabilities actually go down exponentially in $M$.


\begin{theorem}\label{theorem1}
The probabilities of missed detection and false alarm for any user $n$
after $t$ iterations of AMP algorithm, as characterized in
(\ref{eqn:missed detection probability 1}) and
(\ref{eqn:false alarm probability 1})
for the MMSE denoiser based device activity detector,
scale in $M$, the number of receive antennas at the BS, as follows:
\begin{align}\label{eqn:missed detection asymptotic}
P_{t,n}^{\rm MD}(M)
=&-\frac{{\rm exp}(-\frac{1}{2}M\nu_{t,n}^2)}{2\sqrt{2\pi M}}\left(\frac{1}{b_{t,n}-1}+\frac{1}{\nu_{t,n}}\right) \nonumber \\
&\qquad +o\left(\frac{{\rm exp}(-M)}{\sqrt{M}}\right),
\end{align}
and
\begin{align}\label{eqn:false alarm asymptotic}
P_{t,n}^{\rm FA}(M)
=&\frac{{\rm exp}(-\frac{1}{2}M\varsigma_{t,n}^2)}{2\sqrt{2\pi M}}\left(\frac{1}{c_{t,n}-1}+\frac{1}{\varsigma_{t,n}}\right) \nonumber \\
&\qquad +o\left(\frac{{\rm exp}(-M)}{\sqrt{M}}\right),
\end{align}
where $b_{t,n}$ and $c_{t,n}$ are as defined in (\ref{eqn:b1}) and
(\ref{eqn:b2}) respectively, and
\begin{align}
\nu_{t,n}
=& -\sqrt{2(b_{t,n}-1-\log b_{t,n})}, \label{eqn:nu1} \\
\varsigma_{t,n} 
=& \sqrt{2(c_{t,n}-1-\log c_{t,n})}. \label{eqn:nu2}
\end{align}
Further, assuming that $\beta_n$ is bounded below, i.e., $\beta_n >
\beta_{\min}, \forall n$, we then have
$b_{t,n}\leq 1-\varepsilon_{t,n}^{(1)}$,
$c_{t,n}\geq 1+\varepsilon_{t,n}^{(2)}$,
$\nu_{t,n}\leq -\varepsilon_{t,n}^{(3)}$,
$\varsigma_{t,n}\geq \varepsilon_{t,n}^{(4)}$,
for some positive constants $\varepsilon_{t,n}^{(1)}$,
$\varepsilon_{t,n}^{(2)}$, $\varepsilon_{t,n}^{(3)}$, and
$\varepsilon_{t,n}^{(4)}$ that are independent of $M$, and that
\begin{align}
\lim\limits_{M\rightarrow \infty}P_{t,n}^{\rm MD}(M)=\lim\limits_{M\rightarrow \infty}P_{t,n}^{\rm FA}(M)=0, ~~~ \forall t,n.
\end{align}
\end{theorem}

\begin{IEEEproof}
Please refer to Appendix \ref{appendix1}.
\end{IEEEproof}



It is interesting to note that Theorem \ref{theorem1} shows that the
device detection error probability goes to zero as $M \rightarrow
\infty$ for any arbitrary $t$. Thus remarkably, this is true even for
$t=1$, i.e., with infinitely large $M$, detection is already correct
after just one iteration. Further, it is also true for when
$t \rightarrow \infty$.


Theorem \ref{theorem1} is also striking in stating that accurate user
activity detection is guaranteed as long as $M$ is large, regardless
of the relative ratios between $N, K, L$. Thus, this is true even if
$L \le K$. However as shown later in the paper, channel estimation
performance would be poor in this $L \le K$ case.

\subsection{State Evolution in the Asymptotic Massive MIMO Regime}

Theorem \ref{theorem1} states that when $M\rightarrow \infty$, the detection
strategy (\ref{eqn:detection function}) is almost always successful. This also
implies that the MMSE denoiser in the AMP algorithm (\ref{eqn:threshold})
must converge as follows
\begin{align}\label{eqn:detection function 1}
\phi_{t,n} \rightarrow \left\{\begin{array}{ll}1, & {\rm if} ~ \alpha_n=1, \\ 0, & {\rm if} ~ \alpha_n=0. \end{array}\right.\end{align}
Using this fact, the state evolution equation of the AMP algorithm can
be significantly simplified.
Intuitively, since $\phi_{t,n}$ almost surely converges to $0$ or $1$
in the massive MIMO regime, the term $\vartheta_{t,\beta_n}(M)$ in
(\ref{eqn:vartheta 1}), which captures the cost of imperfect user
activity detection, becomes negligible when $M$ goes to infinity.

\begin{theorem}\label{theorem12}
For any user $n$ and at $t$th iteration, as the number of antennas at the BS
$M$ goes to infinity, we have
\begin{align}
\lim\limits_{M\rightarrow \infty}\vartheta_{t,\beta}(M)\rightarrow 0.
\end{align}
\end{theorem}
\begin{IEEEproof}
Please refer to Appendix \ref{appendix6}.
\end{IEEEproof}

As a consequence, as $M$ goes to infinity, the scalar form of the state evolution
for the MMSE denoiser based AMP algorithm given in (\ref{eqn:state evolution fixed M}) reduces to
\begin{align}\label{eqn:state evolution scalar form}
\tau_{t+1}^2=\frac{\sigma^2}{\xi } +
\omega \epsilon \mathbb{E}_{\beta}\left[\frac{\beta\tau_t^2}{\beta+\tau_t^2}\right],
\end{align}
in the massive MIMO regime. 
It is reasonable to deduce that
the fixed-point solution to the state evolution given in
(\ref{eqn:state evolution fixed M}) approaches the fixed-point
solution to the simplified state evolution shown in
(\ref{eqn:state evolution scalar form}).


\subsection{Channel Estimation in Asymptotic Massive MIMO Regime}

As another consequence of (\ref{eqn:detection function 1}),
the covariance matrices of the estimated channels and the
corresponding channel estimation errors as given in
(\ref{eqn:channel 1})-(\ref{eqn:channel 2}) can be considerably
simplified in the massive MIMO regime when $M\rightarrow \infty$.

\begin{theorem}\label{theorem6}
Assuming simplified state evolution (\ref{eqn:state evolution scalar form})
in the massive MIMO regime as $M$, the number of antennas at the BS,
goes to infinity, for each active user $k\in \mathcal{K}$ and after
the $t$th iteration, the covariance matrices of the estimated channel
and channel estimation error are as given in (\ref{eqn:channel 1}) and
(\ref{eqn:channel 2}), where $\upsilon_{t,k}(M)$ and
$\Delta \upsilon_{t,k}(M)$ respectively converge to
\begin{align}
& \lim\limits_{M\rightarrow \infty} \upsilon_{t,k}(M)
	= \frac{\beta_k^2}{\beta_k+\tau_t^2}, \label{eqn:channel_estimate} \\
& \lim\limits_{M\rightarrow \infty} \Delta \upsilon_{t,k}(M)
	=\frac{\beta_k\tau_t^2}{\beta_k+\tau_t^2}.
\label{eqn:channel_error}
\end{align}
\end{theorem}

\begin{IEEEproof}
Please refer to Appendix \ref{appendix7}.
\end{IEEEproof}

This result enables us to characterize the achievable rates of the massive
connectivity system in the massive MIMO regime in Part II of this paper
\cite{LiuPart2}.

\section{Numerical Examples}\label{sec:Numerical Examples}

In this section, we provide numerical examples to verify the results
of this paper. The setup is as follows. There are $N=2,000$ devices
in the cell, and at any time slot, each user accesses the channel with a probability of
$\epsilon=0.05$.
Let $d_n$ denote the distance between user $n$ and the BS, $\forall n$.
It is assumed that $d_n$'s are randomly distributed in the regime
$[0.05{\rm km},1{\rm km}]$.  The path loss model of the wireless
channel for user $n$ is given as $\beta_n=-128.1-36.7\log_{10}(d_n)$
in dB, $\forall n$.  The bandwidth and the coherence time of the
wireless channel are $1$MHz and $1$ms, respectively, and thus in each
coherence block $T=1000$ symbols can be transmitted. The transmit
power for each user at both the first and second transmission phases
is $\rho^{\rm pilot}=23$dBm. The power spectral density of the AWGN at
the BS is assumed to be $-169$dBm/Hz.  Moreover, the numerical results
here are obtained by averaging over $10^7$ channel realizations.

\subsection{Probabilities of Missed Detection and False Alarm}

\begin{figure}
\begin{center}
\scalebox{0.6}{\includegraphics*{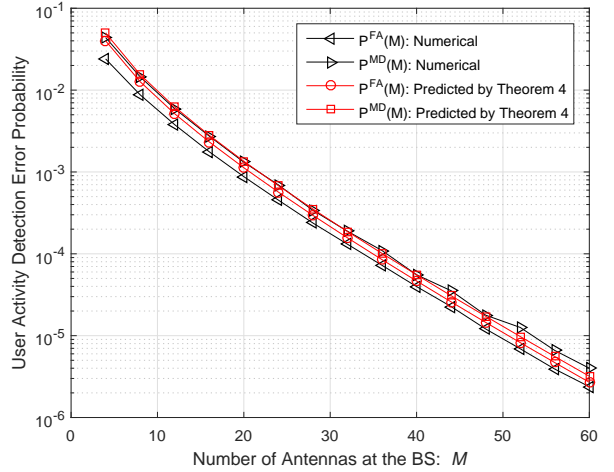}}
\end{center}
\caption{Numerical simulations versus analytical expressions of probabilities of missed
detection and false alarm as functions of the number of antennas
at the BS when each of the $N=2000$ users
accesses the channel with probability $\epsilon=0.05$ at each
coherence time. The transmit power of each user is $\rho^{\rm pilot} =23$dBm.
The pilot sequence length is $L=90$.}\label{fig9}
\end{figure}

First, we examine the performance of user activity detection achieved
by the proposed AMP-based detector. 
Given any $M$, define the average probabilities of missed detection
and false alarm over all users as
$P^{\rm MD}(M)=\sum_{n=1}^NP^{\rm MD}_{\infty,n}(M)/N$ and
$P^{\rm FA}(M)=\sum_{n=1}^NP^{\rm FA}_{\infty,n}(M)/N$, respectively, where
$P^{\rm MD}_{\infty,n}(M)$ and $P^{\rm FA}_{\infty,n}(M)$ denote
the probabilities of missed detection and false alarm of user $n$
after the convergence of the AMP algorithm as defined in
(\ref{eqn:missed detection probability}) and (\ref{eqn:false alarm probability}).
Fig.~\ref{fig9} shows $P^{\rm MD}(M)$ and $P^{\rm FA}(M)$ versus $M$,
the number of antennas at the BS, when the length of the user pilot sequences is $L=90$
both in simulation and as predicted by Theorem \ref{theorem1}.
First, it is observed that the probabilities of missed detection and
false alarm characterized in Theorem \ref{theorem1} match the
numerical results from AMP algorithm very well. Next, it is observed
that as $M$ increases, both $P^{\rm MD}(M)$ and $P^{\rm FA}(M)$
decrease exponentially fast towards zero as predicated by Theorem
\ref{theorem1}.

\begin{figure}
\begin{center}
\scalebox{0.6}{\includegraphics*{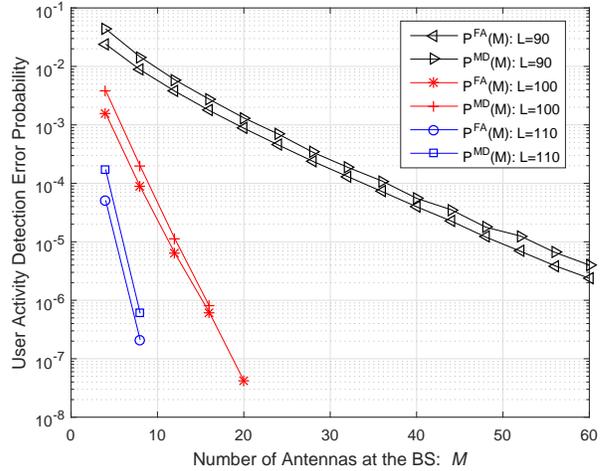}}
\end{center}
\caption{Probabilities of missed detection and false alarm as
functions of the number BS antennas when each of the $N=2000$ users
accesses the channel with probability $\epsilon=0.05$ at each coherence time. The transmit power
of the users is $\rho^{\rm pilot} =23$dBm.}\label{fig3}
\end{figure}

Fig.~\ref{fig3} shows the probabilities of missed detection and
false alarm as functions of $M$ for various values of $L$. It is
observed that although for any $L$, both $P^{\rm MD}(M)$ and
$P^{\rm FA}(M)$ decrease over $M$, but the reduction is more
rapid when $L$ is $110$ as compared to $90$.
Specifically, when $L=90<K$, $M=52$ antennas are needed to drive
$P^{\rm MD}(M)$ and $P^{\rm FA}(M)$ below $10^{-5}$; when
$L=110>K$, only $M=8$ antennas are sufficient. The point is that
although Theorem \ref{theorem1} holds for all $L$ as long as $M
\rightarrow \infty$, in practice the pilot length $L$ should be
chosen to be larger than $K\approx N \epsilon =100$, as otherwise very large number of
antennas would be needed. 

\begin{figure}[t]
\begin{center}
\scalebox{0.6}{\includegraphics*{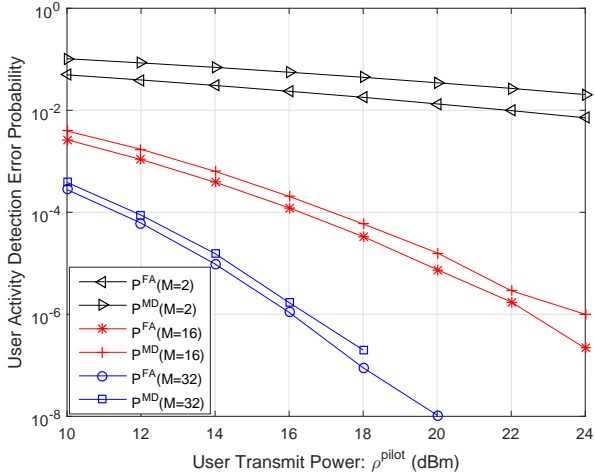}}
\end{center}
\caption{Probabilities of missed detection and false alarm as
functions of the identical transmit power of
all the active users with different numbers of antennas at the BS when each of the $N=2000$ users
accesses the channel with probability $\epsilon=0.05$
at each coherence time and the pilot sequence length is $L=N\epsilon =100$.}\label{fig7}
\end{figure}

Fig.~\ref{fig7} shows the probabilities of missed detection and false
alarm versus the transmit power of the users $\rho^{\rm pilot}$, with
different numbers of antennas at the BS, where the length of the user
pilot sequences is set to be $L=100$. It is observed that for all values of $\rho^{\rm
pilot} $, when there are $M=32$ antennas at the BS, both $P^{\rm
MD}$ and $P^{\rm FA}$ are significantly lower than those in the
cases when there are $M=2$ and $M=16$ antennas at the BS. Further,
when $M=32$, $P^{\rm MD}$ and $P^{\rm FA}$ decrease much faster
with $\rho^{\rm pilot} $ than the cases of $M=2$ and $M=16$.

\begin{figure}[t]
\begin{center}
\scalebox{0.6}{\includegraphics*{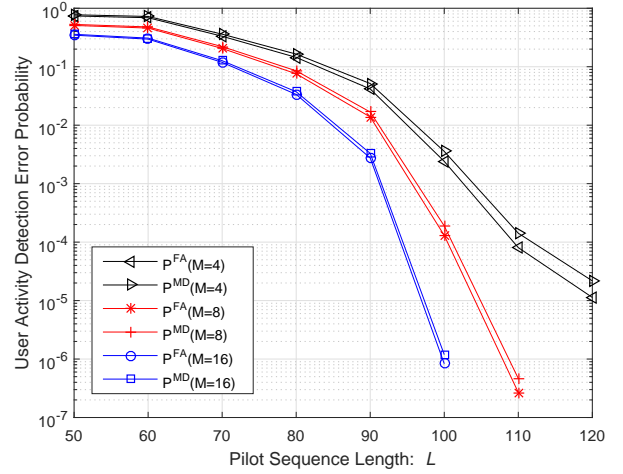}}
\end{center}
\caption{Probabilities of missed detection and false alarm as
functions of the pilot sequence lengths when each of the $N=2000$ users
accesses the channel with probability $\epsilon=0.05$ at each coherence time and the transmit
power of users is $\rho^{\rm pilot} =23$dBm. Further, the BS is equipped with $M=4,8$ or $16$ antennas.}\label{fig8}
\end{figure}

Fig.~\ref{fig8} shows the probabilities of missed detection
and false alarm versus the length of the pilot sequences, $L$, with
different number of antennas at the BS, where the transmit power of each user is
$23$dBm, and the number of antennas at the BS is $M=4,8$ or $16$.
It is observed that both $P^{\rm MD}$ and $P^{\rm FA}$
decrease as the pilot sequence length $L$ increases and when $M$ increases
from $4$ to $8$ and $16$.

\subsection{Channel Estimation Error}

%

\begin{figure}[t]
\begin{center}
\scalebox{0.6}{\includegraphics*{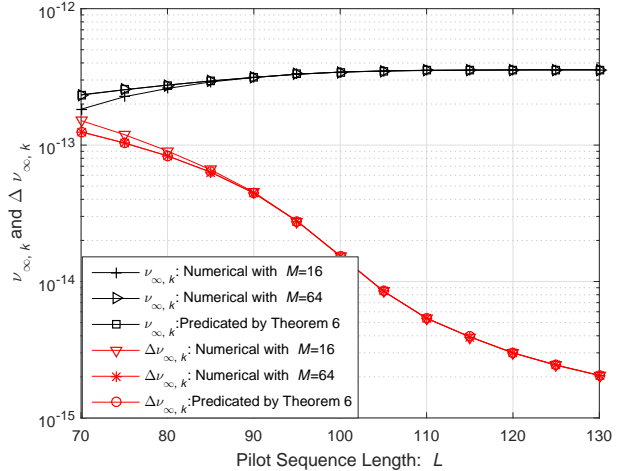}}
\end{center}
\caption{Variance of the estimated channel and the channel estimation
error for a particular user at $0.8$km from the BS as function of pilot
sequence length. Here, each of the $N=2000$ users
accesses the channel with probability $\epsilon=0.05$ in each coherence time; the BS has $M=16$ or $64$ antennas.}\label{fig11}
\end{figure}

Fig.~\ref{fig11} verifies the channel estimation error as predicted by
Theorem \ref{theorem6} after the convergence of the AMP algorithm. In
particular, we select an active user $k$ that is $0.8$km away from the
BS and run AMP to calculate the covariance matrices of its estimated
channel and corresponding channel estimation error numerically, for
two cases of when the BS has $M=16$ or $M=64$ antennas. It is observed
that when $M=64$, both $\nu_{\infty,k}$ and $\Delta \nu_{\infty,k}$
obtained numerically from the AMP algorithm perfectly match those
predicated by Theorem \ref{theorem6} for all values of $L$. When
$M=16$, there is some mismatch in the regime of $L<90$ as device
detection is not perfect in the regime where $M$ and $L$ are both small.
We remark that to have reasonable channel estimation error, the AMP should
operate in the regime of $L>K\approx N\epsilon$.

\section{Conclusion}\label{sec:Conclusion}
Device activity detection and channel estimation are crucial
components for wireless massive connectivity applications. This paper
shows that the vector AMP algorithm is a natural tool for sparse
activity detection.  Utilizing the state evolution analysis, both the
missed detection and false alarm probabilities and channel estimation
error can be analytically characterized.  Further, in the massive MIMO regime,
perfect device activity detection can be guaranteed, but channel estimation
error remains. Part II of this paper \cite{LiuPart2} utilizes these results to further
characterize the achievale rate.

\begin{appendix}

\subsection{Derivation of MMSE Denoiser (\ref{eqn:MMSE denoiser})}
\label{app:lemmas}

\begin{lemma} \label{denoiser_lemma}
Let $\hat{\mv{X}} = \mv{X} + \mv{\Sigma}^{\frac{1}{2}} \mv{V}$
where $\mv{X} \in \mathbb{C}^{M\times 1}$ has a Bernoulli-Gaussian
distribution $(1-\epsilon)\delta_0+\epsilon p_{\mv{h}_\beta}$,
with $\mv{h}_\beta \sim \mathcal{CN}(\mv{0},\beta\mv{I})$,
$\mv{\Sigma}$ is some positive definite matrix and
$\mv{V}\in \mathbb{C}^{M\times 1} \sim \mathcal{CN}(\mv{0},\mv{I})$
is independent of $\mv{X}$.
Define
\begin{align}
&\noindent
\phi(\hat{\mv{x}})= \nonumber \\
&\frac{1}{1+\frac{1-\epsilon}{\epsilon}{\rm exp} \left(-
\hat{\mv{x}}^H\left(\mv{\Sigma}^{-1}-\left(\mv{\Sigma}+\beta\mv{I}\right)^{-1}\right)\hat{\mv{x}} \right)
\left|\mv{I}+\beta\mv{\Sigma}^{-1}\right|}.
\end{align}
Then, 
\begin{align}
\mathbb{E}[\mv{X}|\hat{\mv{X}}=\hat{\mv{x}}]=\phi(\hat{\mv{x}})
	\beta(\beta\mv{I}+\mv{\Sigma})^{-1}\hat{\mv{x}},
\end{align}and
\begin{multline}
\mathbb{E}[\mv{X}\mv{X}^H|\hat{\mv{X}}=\hat{\mv{x}}]
	= \phi(\hat{\mv{x}})(\beta\mv{I}-\beta^2(\beta\mv{I}+\mv{\Sigma})^{-1}
\\ +\beta^2(\beta\mv{I}+\mv{\Sigma})^{-1}\hat{\mv{x}}\hat{\mv{x}}^H(\beta\mv{I}+\mv{\Sigma})^{-1}).
\end{multline}
\end{lemma}

\begin{IEEEproof}
For convenience, define $\bar{\mv{X}} = \mv{h}_\beta + \mv{\Sigma}^{\frac{1}{2}} \mv{V}$. 
Let $\hat{\mv{X}}=\bar{\mv{X}}$ with probability $\epsilon$
and $\hat{\mv{X}}={\mv{\Sigma}^{\frac{1}{2}}\mv{V}}$ with probability $1-\epsilon$. By standard estimation theory, we have
\begin{align}
\mathbb{E}[\mv{h}_\beta|\bar{\mv{X}}=\bar{\mv{x}}]=
	\beta(\beta\mv{I}+\mv{\Sigma})^{-1}\bar{\mv{x}},
\end{align}and
\begin{multline}
\mathbb{E}[\mv{h}_\beta\mv{h}_\beta^H|\bar{\mv{X}}=\bar{\mv{x}}]
	= \beta\mv{I}-\beta^2(\beta\mv{I}+\mv{\Sigma})^{-1}
\\ +\beta^2(\beta\mv{I}+\mv{\Sigma})^{-1}\bar{\mv{x}}\bar{\mv{x}}^H(\beta\mv{I}+\mv{\Sigma})^{-1}.
\end{multline}
We now characterize $\mathbb{E}[\mv{X}|\hat{\mv{X}}=\hat{\mv{x}}]$ and $\mathbb{E}[\mv{X}\mv{X}^H|\hat{\mv{X}}=\hat{\mv{x}}]$
based on $\mathbb{E}[\mv{h}_\beta| \bar{\mv{X}}=\bar{\mv{x}}]$ and $\mathbb{E}[\mv{h}_\beta\mv{h}_\beta^H|\bar{\mv{X}}=\bar{\mv{x}}]$.
%
\begin{align}
& \noindent \mathbb{E}[\mv{X}|\hat{\mv{X}}=\hat{\mv{x}}]
 = \int \mv{x}p_{\mv{X}|\hat{\mv{X}}}(\mv{X}=\mv{x}|\hat{\mv{X}}=\hat{\mv{x}}) d\mv{x} \\
& = \frac{1}{p_{\hat{\mv{X}}}(\hat{\mv{X}}=\hat{\mv{x}})}\int \mv{x}
p_{\hat{\mv{X}}|\mv{X}}(\hat{\mv{X}}=\hat{\mv{x}}|\mv{X}=\mv{x})
\nonumber \\
& \qquad \qquad \qquad \left( \epsilon p_{\mv{h}_\beta}(\mv{h}_\beta=\mv{x}) +
	(1-\epsilon) \delta_0(\mv{x}) \right) d\mv{x} \\
& = \frac{\epsilon}{p_{\hat{\mv{X}}}(\hat{\mv{X}}=\hat{\mv{x}})}\int \mv{x} p_{\bar{\mv{X}},\mv{h}_\beta}(\bar{\mv{X}}=\hat{\mv{x}}|\mv{h}_\beta=\mv{x}) \nonumber \\
& \qquad \qquad \qquad \qquad \qquad \qquad \qquad p_{\mv{h}_\beta}(\mv{h}_\beta=\mv{x}) d\mv{x} \label{eqn:lemma 1} \\
& = \frac{\epsilon p_{\bar{\mv{X}}}(\bar{\mv{X}}=\hat{\mv{x}})}{p_{\hat{\mv{X}}}(\hat{\mv{X}}=\hat{\mv{x}})}\mathbb{E}[\mv{h}_\beta|\bar{\mv{X}}=\hat{\mv{x}}] \\
& = \phi(\hat{\mv{x}})\beta(\beta\mv{I}+\mv{\Sigma})^{-1}\hat{\mv{x}}, \label{eqn:lemma 2}
\end{align}
where (\ref{eqn:lemma 1}) is because given $\mv{X}=\mv{x}$,
$p_{\hat{\mv{X}}|{\mv{X}}}(\hat{\mv{X}} = \hat{\mv{x}}| {\mv{X}} = {\mv{x}})$
is just the distribution of the random vector $\mv{x}+\mv{\Sigma}^{\frac{1}{2}}\mv{V}$, but
the same is true for 
$p_{\bar{\mv{X}}|{\mv{h}_\beta}}(\bar{\mv{X}} = \hat{\mv{x}}| {\mv{h}_\beta} = {\mv{x}})$,
and (\ref{eqn:lemma 2}) is because
$p_{\hat{\mv{X}}}(\hat{\mv{X}} =\hat{\mv{x}})$ takes the form of
%
\begin{align}
\epsilon \frac{{\rm exp}(-\hat{\mv{x}}^H(\beta\mv{I}+\mv{\Sigma})^{-1}\hat{\mv{x}})}{\pi^M|\beta\mv{I}+\mv{\Sigma}|}+ (1-\epsilon) \frac{{\rm exp}(-\hat{\mv{x}}^H\mv{\Sigma}^{-1}\hat{\mv{x}})}{\pi^M|\mv{\Sigma}|} \label{eqn:lemma probability}
\end{align}
so a straightforward computation gives that
\begin{align}\label{eqn:lemma 3}
\frac{\epsilon p_{\bar{\mv{X}}}(\bar{\mv{X}}=\hat{\mv{x}})}{p_{\hat{\mv{X}}}(\hat{\mv{X}}=\hat{\mv{x}})}=\phi(\hat{\mv{x}}).
\end{align}
Similarly, it can be shown that $\mathbb{E}[\mv{X}\mv{X}^H|\hat{\mv{X}}=\hat{\mv{x}}]
	= \phi(\hat{\mv{x}})\mathbb{E}[\mv{h}_\beta\mv{h}_\beta^H|\bar{\mv{X}}=\hat{\mv{x}}]$. 
\end{IEEEproof}

\subsection{Proof of Theorem \ref{theorem7}}\label{appendix11}

We first characterize the MSE term in
(\ref{eqn:state evolution}) when the MMSE denoiser (\ref{eqn:MMSE
denoiser}) is used in the vector AMP algorithm, then show that the
state $\mv{\Sigma}_t$ stays as a diagonal matrix with identical
diagonal entries throughout the AMP iterations.

Let $\hat{\mv{X}}_{t,\beta} = \mv{X}_\beta + \Sigma_t^{\frac{1}{2}} \mv{V}$.
For a fixed $\beta$, the expectation term in (\ref{eqn:state evolution}) with
the MMSE denoiser $\eta_{t,\beta}(\hat{\mv{x}}_{t,\beta}) =
\mathbb{E}[\mv{X}_\beta|\hat{\mv{X}}_{t,\beta}=\hat{\mv{x}}_{t,\beta}]$
can be expressed as \cite{zhilin_JSAC}
\begin{align}
& \noindent \mathbb{E}_{\hat{\mv{X}}_{t,\beta} \mv{X}_\beta}
\bigg[\left(\eta_{t,\beta}(\hat{\mv{X}}_{t,\beta})-\mv{X}_\beta)(\eta_{t,\beta}(\hat{\mv{X}}_{t,\beta})-\mv{X}_\beta\right)^H\bigg]  \nonumber \\
& = \mathbb{E}_{\hat{\mv{X}}_{t,\beta}}
\mathbb{E}_{\mv{X}_\beta|\hat{\mv{X}}_{t,\beta}=\hat{\mv{x}}_{t,\beta}}
\bigg[\left(\mathbb{E}[\mv{X}_\beta|\hat{\mv{X}}_{t,\beta}=\hat{\mv{x}}_{t,\beta}]-\mv{X}_\beta\right) \nonumber \\
& \qquad \qquad \qquad \qquad \left(\mathbb{E}[\mv{X}_\beta|\hat{\mv{X}}_{t,\beta}=\hat{\mv{x}}_{t,\beta}]-\mv{X}_\beta\right)^H\bigg] \nonumber \\
& = \mathbb{E}_{\hat{\mv{X}}_{t,\beta}}
\mathbb{E}\left[\mv{X}_\beta\mv{X}_\beta^H \left|\hat{\mv{X}}_{t,\beta}=\hat{\mv{x}}_{t,\beta}\right.\right]
\nonumber \\
& \qquad - \mathbb{E}_{\hat{\mv{X}}_{t,\beta}}
\mathbb{E}\left[\mv{X}_\beta \left|\hat{\mv{X}}_{t,\beta}=\hat{\mv{x}}_{t,\beta}\right.\right](\mathbb{E}[\mv{X}_\beta|\hat{\mv{X}}_{t,\beta}=\hat{\mv{x}}_{t,\beta}])^H \nonumber \\
& = \mathbb{E}_{\hat{\mv{X}}_{t,\beta}}[\phi_{t,\beta}(\beta\mv{I}-\beta^2(\beta\mv{I}+\mv{\Sigma}_t)^{-1})+ \nonumber
\\ & \qquad \phi_{t,\beta}(1-\phi_{t,\beta})\beta^2(\beta\mv{I} + \mv{\Sigma}_t)^{-1}\hat{\mv{x}}_{t,\beta}\hat{\mv{x}}_{t,\beta}^H(\beta \mv{I} + \mv{\Sigma}_t)^{-1}], \label{eqn:MSE appendix}
\end{align}
where the last equality is due to Lemma \ref{denoiser_lemma} and
$\phi_{t,\beta}$ is used to denote $\phi(\hat{\mv{x}}_{t,\beta})$.

According to (\ref{eqn:lemma 3}), we have
\begin{align}\label{eqn:Gaussian distribution}
\phi_{t,\beta}p_{\hat{\mv{x}}_{t,\beta}}=\frac{\epsilon {\rm exp}(-\hat{\mv{x}}_{t,\beta}^H(\beta\mv{I}+\mv{\Sigma}_t)^{-1}\hat{\mv{x}}_{t,\beta})}{\pi^M|\beta\mv{I}+\mv{\Sigma}_t|}.
\end{align}
As a result, we have $\mathbb{E}_{\hat{\mv{X}}_{t,\beta}}[\phi_{t,\beta}]=\epsilon$.
It then follows that
\begin{multline}
\mathbb{E}_{\hat{\mv{X}}_{t,\beta}}\left[\phi_{t,\beta}(\beta\mv{I}-\beta^2(\beta\mv{I}+\mv{\Sigma}_t)^{-1})\right] \\
= \epsilon (\beta\mv{I}-\beta^2(\beta\mv{I}+\mv{\Sigma}_t)^{-1}). \label{eqn:expectation of phi}
\end{multline}

By substituting (\ref{eqn:expectation of phi}) into (\ref{eqn:MSE appendix}),
the expected MSE for estimating $\mv{X}_\beta$ given any $\beta$ after
$t$ iterations is
\begin{multline}\label{eqn:MSE for AMP}
\mathbb{E}_{\hat{\mv{X}}_{t,\beta}}\left[(\eta_{t,\beta}(\hat{\mv{X}}_{t,\beta})-\mv{X}_\beta)(\eta_{t,\beta}(\hat{\mv{X}}_{t,\beta})-\mv{X}_\beta)^H\right] \\
= \epsilon(\beta\mv{I}-\beta^2(\beta\mv{I}+\mv{\Sigma}_t)^{-1}) +
\mathbb{E}_{\hat{\mv{X}}_{t,\beta}}[\phi_{t,\beta}(1-\phi_{t,\beta})\beta^2
\\ (\beta\mv{I} + \mv{\Sigma}_t)^{-1} \hat{\mv{X}}_{t,\beta}\hat{\mv{X}}_{t,\beta}^H(\beta\mv{I} + \mv{\Sigma}_t)^{-1}].
\end{multline}
Note here $\phi_{t,\beta}$ is also a random variable depending on $\hat{\mv{X}}_{t,\beta}$ as in (\ref{eqn:threshold}).


Based on the MSE shown in (\ref{eqn:MSE for AMP}), in the following we
show that $\mv{\Sigma}_t$ stays as a diagonal matrix with identical
diagonal entries after each iteration of the AMP algorithm.

First, it can be easily seen that for the initial state given in (\ref{eqn:state evolution initial point}),
$\mv{\Sigma}_t$ is indeed a diagonal matrix with identical diagonal elements, i.e., $\mv{\Sigma}_0=\tau_0^2\mv{I}$, with
\begin{align}
\tau_0^2=\frac{\sigma^2}{\xi}+\omega\epsilon\mathbb{E}[\beta].
\end{align}

Next, suppose that $\mv{\Sigma}_t=\tau_t^2\mv{I}$, given $t\geq 0$. We show in the following that
$\mv{\Sigma}_{t+1}$ must be a diagonal matrix with identical diagonal elements, i.e., $\mv{\Sigma}_{t+1}=\tau_{t+1}^2\mv{I}$. Define
\begin{align}
\mv{D} & = \mathbb{E}[\phi_{t,\beta}(1-\phi_{t,\beta})\beta^2(\beta\mv{I} + \mv{\Sigma}_t)^{-1}\hat{\mv{X}}_{t,\beta}\hat{\mv{X}}_{t,\beta}^H \nonumber \\
& \qquad \qquad \qquad \qquad \qquad \qquad \qquad (\beta\mv{I} + \mv{\Sigma}_t)^{-1}] \nonumber \\
& = \mathbb{E}\left[\phi_{t,\beta}(1-\phi_{t,\beta})\frac{\beta^2}{(\beta + \tau_t^2)^{2}}\hat{\mv{X}}_{t,\beta}\hat{\mv{X}}_{t,\beta}^H\right],
\end{align}
where the distribution of $\hat{\mv{X}}_{t,\beta}$ given in (\ref{eqn:lemma probability}) reduces to
\begin{align}\label{eqn:distribution hat x}
p_{\hat{\mv{x}}_{t,\beta}} = 
\frac{\epsilon {\rm exp}(-\frac{\hat{\mv{x}}_{t,\beta}^H\hat{\mv{x}}_{t,\beta}}{\beta+\tau_t^2})}{\pi^M(\beta+\tau_t^2)^M}+\frac{(1-\epsilon) {\rm exp}(-\frac{\hat{\mv{x}}_{t,\beta}^H\hat{\mv{x}}_{t,\beta}}{\tau_t^2})}{\pi^M(\tau_t^2)^M},
\end{align}
and the random variable $\phi_{t,\beta}$ given in (\ref{eqn:threshold}) reduces to
\begin{align}\label{eqn:threshold new}
\phi_{t,\beta}=\frac{1}{1+\frac{1-\epsilon}{\epsilon}{\rm
exp}\left(-\left(\frac{1}{\tau_t^2}-\frac{1}{\beta+\tau_t^2}\right)\hat{\mv{x}}_{t,\beta}^H\hat{\mv{x}}_{t,\beta}\right)\left(1+\frac{\beta}{\tau_t^2}\right)^M}.
\end{align}
For any $1\leq i,j\leq M$, define $\mv{D}(i,j)$ as the element in the $i$th row and $j$th column
of $\mv{D}$, and $\hat{\mv{X}}_{t,\beta}(i)$ as the $i$th element of $\hat{\mv{X}}_{t,\beta}$. For the non-diagonal elements of $\mv{D}$, we have
\begin{align}
& \mv{D}(i,j)\nonumber \\
& =\mathbb{E}[\phi_{t,\beta}(1-\phi_{t,\beta})\frac{\beta^2}{(\beta + \tau_t^2)^{2}}\hat{\mv{X}}_{t,\beta}(i)\hat{\mv{X}}_{t,\beta}^H(j)] \\
& =\int \phi_{t,\beta}(1-\phi_{t,\beta})\frac{\beta^2}{(\beta + \tau_t^2)^{2}}\hat{\mv{x}}_{t,\beta}(i)\hat{\mv{x}}_{t,\beta}^H(j) p_{\hat{\mv{x}}_{t,\beta}} d \hat{\mv{x}}_{t,\beta}, 
\end{align}
where $\hat{\mv{x}}_{t,\beta}(i)$ is
the $i$th element of $\hat{\mv{x}}_{t,\beta}$. The key observation is
that $\phi_{t,\beta}$ as expressed in (\ref{eqn:threshold new}) and
$p_{\hat{\mv{x}}_{t,\beta}}$ as expressed in (\ref{eqn:distribution hat x})
involves only square terms $|\hat{\mv{x}}_{t,\beta}(i)|^2$ so are both
even functions, while
$\hat{\mv{x}}_{t,\beta}(i)\hat{\mv{x}}_{t,\beta}^H(j)$ is an odd
function for $i \neq j$, so the overall integral is zero.
Hence $\mv{D}(i,j)=0$ for $i \neq j$, i.e., $\mv{D}$ is diagonal.

Next, consider the diagonal terms of $\mv{D}$
\begin{align}\label{eqn:vartheta}
\mv{D}(i,i)=\mathbb{E}[\phi_{t,\beta}(1-\phi_{t,\beta})\frac{\beta^2}{(\beta + \tau_t^2)^{2}}\hat{\mv{X}}_{t,\beta}(i)\hat{\mv{X}}_{t,\beta}^H(i)].
\end{align}It can be observed from (\ref{eqn:distribution hat x}) and (\ref{eqn:threshold new}) that all the elements of $\hat{\mv{X}}_{t,\beta}$ contribute
equally to $p_{\hat{\mv{x}}_{t,\beta}}$ and $\phi_{t,\beta}$. Moreover,
all the elements of $\hat{\mv{X}}_{t,\beta}$ have an identical distribution. As a result, we have $\mv{D}(i,i)=\mv{D}(j,j)=\vartheta_{t,\beta}(\tau_t^2)$, $1\leq i,j\leq M$, i.e., $\mv{D}=\vartheta_{t,\beta}(\tau_t^2)\mv{I}$.
To compute $\vartheta_{t,\beta}(\tau_t^2)$, we observe that
\begin{align}
\vartheta_{t,\beta}(\tau_t^2)=\frac{{\rm tr}(\mv{D})}{M}=\frac{\mathbb{E}[\phi_{t,\beta}(1-\phi_{t,\beta})\frac{\beta^2}{(\beta+\tau_t)^2}\hat{\mv{X}}_{t,\beta}^H\hat{\mv{X}}_{t,\beta}]}{M}.
\end{align}
This proves (\ref{eqn:vartheta 1}).

Finally, putting the above expression for $\mv{D}$ back to
(\ref{eqn:MSE for AMP}) and recognizing that
if $\mv{\Sigma}_t=\tau_t^2\mv{I}$, then
\begin{align}
\epsilon(\beta\mv{I}-\beta_n^2(\beta_n\mv{I}+\mv{\Sigma}_t)^{-1})=\epsilon \left(\frac{\beta\tau_t^2}{\beta+\tau_t^2}\mv{I}\right),
\end{align}
so by the state evolution (\ref{eqn:state evolution}), this implies
that if $\mv{\Sigma}_t=\tau_t^2\mv{I}$, we have
\begin{align}
\mv{\Sigma}_{t+1}=\frac{\sigma^2}{\xi}\mv{I}+\omega \left(\epsilon \mathbb{E}\left[\frac{\beta\tau_t^2}{\beta+\tau_t^2}\right]\mv{I}+\mathbb{E}[\vartheta_{t,\beta}(\tau_t^2)] \mv{I}\right),
\end{align}
which is also a diagonal matrix with identical diagonal elements, i.e., $\mv{\Sigma}_{t+1}=\tau_{t+1}^2\mv{I}$.
The scalar form of the denoiser (\ref{eqn:MMSE denoiser 1})-(\ref{eqn:psi 2})
and the scalar state evolution (\ref{eqn:state evolution fixed M})
then follow immeidately. Theorem \ref{theorem7} is thus proved.

\subsection{Proof of Theorem \ref{theorem15}}\label{appendix15}

According to the state evolution equivalent signal model
(\ref{eqn:equivalent channel 1}) in Theorem \ref{theorem7},
when $\alpha_n=0$, the entries of $(\mv{R}^t)^H\mv{a}_n+\mv{x}_n^t$ are i.i.d.\
Gaussian distributed with variance $\tau_t^2$.
When $\alpha_n=1$, the entries of $(\mv{R}^t)^H\mv{a}_n+\mv{x}_n^t$ are i.i.d.\
Gaussian distributed with variance $\beta_n+\tau_t^2$.
Since $(\mv{R}^t)^H\mv{a}_n+\mv{x}_n^t$ given $\alpha_n$ is a
complex Gaussian random vector with i.i.d.\ real and imaginary
components, the random variables
$((\mv{R}^t)^H\mv{a}_n+\mv{x}_n^t)^H ((\mv{R}^t)^H\mv{a}_n+\mv{x}_n^t) / (2\tau_t^2)$
given $\alpha_n=0$ and
$((\mv{R}^t)^H\mv{a}_n+\mv{x}_n^t)^H ((\mv{R}^t)^H\mv{a}_n+\mv{x}_n^t) /(2(\beta_n+\tau_t^2))$
given $\alpha_n=1$ thus follow $\chi^2$ distribution with $2M$ DoF.

Let $X \sim \chi_{2M}$ be a random variable that follows $\chi^2$
distribution with $2M$ DoF. It is well known that its cumulative
distribution function follows ${\rm Pr}(X \leq x)=
\frac{\underline{\Gamma}\left(M,\frac{x}{2}\right)}{\Gamma(M)}$.
The proposed detector (\ref{eqn:detection function}) compares
$((\mv{R}^t)^H\mv{a}_n+\mv{x}_n^t)^H ((\mv{R}^t)^H\mv{a}_n+\mv{x}_n^t)$ with the threshold
$M \log\left(1+\frac{\beta_n}{\tau_t^2}\right) /
\left(\frac{1}{\tau_t^2}-\frac{1}{\tau_t^2+\beta_n}\right)$.
By the definition of missed detection and false alarm probabilities
(\ref{eqn:missed detection probability}) and
(\ref{eqn:false alarm probability}), it thus follows that
\begin{align}
P^{\rm MD}_{t,n}(M)=&{\rm Pr}\left( X \leq
\frac{2M \log\left(1+\frac{\beta_n}{\tau_t^2}\right)}{(\beta_n+\tau_t^2)\left(\frac{1}{\tau_t^2}-\frac{1}{\beta_n+\tau_t^2}\right)}\right) \nonumber \\
=&\frac{\underline{\Gamma}\left(M,b_{t,n}M\right)}{\Gamma(M)},
\end{align}and
\begin{align}
P^{\rm FA}_{t,n}(M)=&{\rm Pr}\left(X \geq
\frac{2M \log\left(1+\frac{\beta_n}{\tau_t^2}\right)}
{\tau_t^2\left(\frac{1}{\tau_t^2}-\frac{1}{\beta_n+\tau_t^2}\right)}\right) \nonumber \\
=&1-\frac{\underline{\Gamma}\left(M,c_{t,n}M\right)}{\Gamma(M)} = \frac{\bar{\Gamma}\left(M,c_{t,n}M\right)}{\Gamma(M)},
\end{align}
where
\begin{align}
& b_{t,n}= \frac{
\log\left(1+\frac{\beta_n}{\tau_t^2}\right)
}{(\beta_n+\tau_t^2)\left(\frac{1}{\tau_t^2}-\frac{1}{\beta_n+\tau_t^2}\right)}=
\frac{\tau_t^2}{\beta_n}
\log\left(1+\frac{\beta_n}{\tau_t^2}\right), \\ 
& c_{t,n}= \frac{\log\left(1+\frac{\beta_n}{\tau_t^2}\right)
}{\tau_t^2\left(\frac{1}{\tau_t^2}-\frac{1}{\beta_n+\tau_t^2}\right)}=
\left(\frac{\beta_n + \tau_t^2}{\beta_n} \right)
\log\left(1+\frac{\beta_n}{\tau_t^2}\right).
\end{align}
Theorem \ref{theorem15} is thus proved.

\subsection{Proof of Theorem \ref{theorem9}}\label{appendix9}

Based on the vector AMP state evolution, the estimated channel
and the corresponding channel estimation error, assuming that device $k$
is active, i.e., $\alpha_k=1$, is statistically equivalent to applying
the denoiser (\ref{eqn:MMSE denoiser 1}) to the equivalent signal
model (\ref{eqn:equivalent channel 1}) as shown below:
\begin{align}
& \hat{\mv{h}}_k = 
	\phi_{t,k} \left(\frac{\beta_k}{\beta_k+\tau_\infty^2} \right)
	(\mv{h}_k+\tau_t\mv{v}_k), \label{eqn:hat_h_k} \\
& \Delta \hat{\mv{h}}_k =\mv{h}_k-\hat{\mv{h}}_k, ~~~ \forall k\in \mathcal{K}.
	\label{eqn:delta_h_k}
\end{align}
The main conclusion of Theorem \ref{theorem9} is that
for any active user $k$, given any $M$,
${\rm Cov}(\hat{\mv{h}}_k,\hat{\mv{h}}_k)$ is a diagonal matrix
with identical diagonal entries. The proof of this fact uses
the same technique as used in Appendix \ref{appendix11} for
proving Theorem \ref{theorem7}, and so is not repeated here.
Similarly, it can also be shown that
${\rm Cov}(\Delta \hat{\mv{h}}_k,\Delta \hat{\mv{h}}_k)$ is
a diagonal matrix with identical diagonal entries.
The expressions for the diagonal terms $\upsilon_{t,k}(M)$
and $\Delta \upsilon_{t,k}(M)$
(\ref{eqn:channel 3})-(\ref{eqn:channel 4}) follow directly
from (\ref{eqn:hat_h_k})-(\ref{eqn:delta_h_k}).


\subsection{Proof of Theorem \ref{theorem1}}\label{appendix1}

First we show that assuming $\beta_n$ is bounded below, i.e.,
$\beta_n > \beta_{\min}, \forall n$, we have
$b_{t,n}\leq 1-\varepsilon_{t,n}^{(1)}$,
$c_{t,n}\geq 1+\varepsilon_{t,n}^{(2)}$,
$\nu_{t,n}\leq -\varepsilon_{t,n}^{(3)}$,
$\varsigma_{t,n}\geq \varepsilon_{t,n}^{(4)}$,
for some positive constants $\varepsilon_{t,n}^{(1)}$,
$\varepsilon_{t,n}^{(2)}$, $\varepsilon_{t,n}^{(3)}$, and
$\varepsilon_{t,n}^{(4)}$ that are independent of $M$.
This is because it can be easily checked based on the state
evolution equation (\ref{eqn:state evolution fixed M})
that $\tau_t^2$ is always bounded from above.
So, $\beta_n/\tau_t^2$ is always lower bounded by some positive
constant indepedent of $M$.
Now, combining with the fact that $a > \log(1+a) > \frac{a}{1+a}$
for all $a>0$ with the first inequality becoming equality
if and only if $a=1$, we can see that
$b_{t,n}$, $c_{t,n}$, as in (\ref{eqn:b1})-(\ref{eqn:b2}), and
$\nu_{t,n}$, $\varsigma_{t,n}$ as in (\ref{eqn:nu1})-(\ref{eqn:nu2})
are all bounded away from 1 and 0, respectively, as required.

Next, we study the asymptotic probabilities of missed detection and false alarm.
According to \cite{Gautschi}, if $b_{t,n}\leq
1-\varepsilon_{t,n}^{(1)}$ for some positive constant
$\varepsilon_{t,n}^{(1)}$, then 
\begin{multline}
\frac{\underline{\Gamma}(M,b_{t,n}M)}{\Gamma(M)} = \frac{1}{2}{\rm
efrc}\left(-\nu_{t,n}\sqrt{\frac{M}{2}}\right)
-\frac{{\rm exp}(-\frac{1}{2}M\nu_{t,n}^2)}{\sqrt{2\pi M}} \\
\left(\frac{1}{b_{t,n}-1}-\frac{1}{\nu_{t,n}}\right) -
o\left(\frac{{\rm exp}(-M)}{\sqrt{M}}\right),
\end{multline}
where ${\rm efrc}(\cdot)$ is the complementary error
function. Similarly, if $c_{t,n}\geq 1+\varepsilon_{t,n}^{(2)}$ for
some positive constant $\varepsilon_{t,n}^{(2)}$, then 
\begin{multline}
\frac{\bar{\Gamma}(M,c_{t,n}M)}{\Gamma(M)} = \frac{1}{2}{\rm efrc}\left(\varsigma_{t,n}\sqrt{\frac{M}{2}}\right)+\frac{{\rm exp}(-\frac{1}{2}M\varsigma_{t,n}^2)}{\sqrt{2\pi M}} \\
\left(\frac{1}{c_{t,n}-1}-\frac{1}{\varsigma_{t,n}}\right) + o\left(\frac{{\rm exp}(-M)}{\sqrt{M}}\right).
\end{multline}
Moreover, it is known that 
\begin{align}\label{eqn:asymptotic gamma}
{\rm efrc}(x)=\frac{{\rm exp}(-x^2)}{\sqrt{\pi }x}\left(1+o\left(\frac{1}{x^2}\right)\right).
\end{align}
Then, 
the missed detection probability given in (\ref{eqn:missed detection probability 1}) is
\begin{align}
P_{t,n}^{\rm MD}(M) =&\frac{1}{2}\frac{\underline{\Gamma}(M,b_{t,n}M)}{\Gamma(M)} \nonumber \\
=&\frac{1}{2}\frac{{\rm exp}(-\frac{1}{2}M\nu_{t,n}^2)}{-\nu_{t,n}\sqrt{\pi M/2}}\left(1+o\left(\frac{1}{M}\right)\right) \nonumber \\ & -\frac{1}{2}\frac{{\rm exp}(-\frac{1}{2}M\nu_{t,n}^2)}{\sqrt{2\pi M}}\left(\frac{1}{b_{t,n}-1}-\frac{1}{\nu_{t,n}}\right) \nonumber \\ & -o\left(\frac{{\rm exp}(-M)}{\sqrt{M}}\right) \label{eqn:asymptotic error function}
\\
=&-\frac{{\rm exp}(-\frac{1}{2}M\nu_{t,n}^2)}{2\sqrt{2\pi M}}\left(\frac{1}{b_{t,n}-1}+\frac{1}{\nu_{t,n}}\right) \nonumber \\
&+o\left(\frac{{\rm exp}(-M)}{\sqrt{M}}\right),
\end{align}where (\ref{eqn:asymptotic error function}) is obtained by applying (\ref{eqn:asymptotic gamma}) since $\nu_{t,n}<0$ with $b_{t,n}<1$.
and
the false alarm probability given in (\ref{eqn:false alarm probability 1}) is
\begin{align}
P_{t,n}^{\rm FA}(M) =&\frac{1}{2}\frac{\bar{\Gamma}(M,b_{t,n}M)}{\Gamma(M)} \nonumber \\
=&\frac{1}{2}\frac{{\rm exp}(-\frac{1}{2}M\varsigma_{t,n}^2)}{\varsigma_{t,n}\sqrt{\pi M/2}}\left(1+o\left(\frac{1}{M}\right)\right) \nonumber \\ & +\frac{1}{2}\frac{{\rm exp}(-\frac{1}{2}M\varsigma_{t,n}^2)}{\sqrt{2\pi M}}\left(\frac{1}{c_{t,n}-1}-\frac{1}{\varsigma_{t,n}}\right) \nonumber \\ & +o\left(\frac{{\rm exp}(-M)}{\sqrt{M}}\right)
\nonumber \\
=& \frac{{\rm exp}(-\frac{1}{2}M\varsigma_{t,n}^2)}{2\sqrt{2\pi M}}\left(\frac{1}{c_{t,n}-1}+\frac{1}{\varsigma_{t,n}}\right) \nonumber \\
&+o\left(\frac{{\rm exp}(-M)}{\sqrt{M}}\right).
\end{align}
Theorem \ref{theorem1} is thus proved.

\subsection{Proof of Theorem \ref{theorem12}}\label{appendix6}

Examining the functional form of $\vartheta_{t,\beta}(\tau_t^2)$:
\begin{align}
\vartheta_{t,\beta}(\tau_t^2) =
\mathbb{E}_{\hat{\mv{X}}_{t,\beta}}
\left[\phi_{t,\beta}(1-\phi_{t,\beta})\frac{\beta^2}{(\beta+\tau_t^2)^2}\frac{\hat{\mv{X}}_{t,\beta}^H\hat{\mv{X}}_{t,\beta}}{M}\right],
\end{align}
we see that as $0 \le \phi_{t,\beta} \le 1$ and by the law of large
numbers, the term inside expectation is bounded from above by a
constant independent of $M$.  Thus, we can apply dominated convergence
theorem for taking limit as
$M \rightarrow \infty$, i.e.,
\begin{multline}
\lim_{M \rightarrow \infty} \vartheta_{t,\beta}(\tau_t^2) =
\mathbb{E}_{\hat{\mv{X}}_{t,\beta}}
\left[
\lim_{M \rightarrow \infty}
\phi_{t,\beta}(1-\phi_{t,\beta})\right. \\
\left.
\frac{\beta^2}{(\beta+\tau_t^2)^2}\frac{\hat{\mv{X}}_{t,\beta}^H\hat{\mv{X}}_{t,\beta}}{M}\right].
\end{multline}
But $\lim_{M\rightarrow \infty} \phi_{t,n}$ is either 0 or 1,
according to (\ref{eqn:detection function 1}) as consequence of
Theorem \ref{theorem1}.
It thus follows that
\begin{align}\label{eqn:asymptotic vartheta}
\lim \limits_{M\rightarrow \infty} \vartheta_{t,\beta}(\tau_t^2)=0.
\end{align}

\subsection{Proof of Theorem \ref{theorem6}}\label{appendix7}

Assuming fixed $\tau_t^2$ as given by the simplified state evolution
(\ref{eqn:state evolution scalar form}) in the massive MIMO regime in
the limit as $M$ goes to infinity, we evaluate
$\lim\limits_{M\rightarrow \infty}
\upsilon_k(M)$ and $\lim\limits_{M\rightarrow \infty} \Delta \upsilon_k(M)$
where $\upsilon_k(M)$ and $\Delta \upsilon_k(M)$ are
as given in (\ref{eqn:channel 3}) and (\ref{eqn:channel 4}), respectively.

It is easy to check that since $0\leq \phi_{t,k}^2 \leq 1$ according
to (\ref{eqn:threshold}), it follows that
\begin{align}
\upsilon_k(M) \leq & \mathbb{E}\left[\frac{\beta_k^2}{(\beta_k+\tau_t^2)^2}\frac{(\mv{h}_k+\tau_t\mv{v}_k)^H(\mv{h}_k+\tau_t\mv{v}_k)}{M}\right] \nonumber \\
 = & \frac{\beta_k^2}{\beta_k+\tau_t^2}.
\end{align}
So $\upsilon_k(M)$ is upper bounded by a finite constant indepedent of $M$,
we can then apply the dominated convergence theorem to compute
\begin{align}
& \noindent \lim\limits_{M\rightarrow t} \upsilon_k(M) \nonumber \\
& = \mathbb{E}\left[\lim\limits_{M\rightarrow t} \frac{\phi_{t,k}^2\beta_k^2}{(\beta_k+\tau_t^2)^2}\frac{(\mv{h}_k+\tau_t\mv{v}_k)^H(\mv{h}_k+\tau_t\mv{v}_k)}{M}\right] \\
& = \mathbb{E}\left[\frac{\beta_k^2}{(\beta_k+\tau_t^2)^2} \lim\limits_{M\rightarrow t} \frac{(\mv{h}_k+\tau_t\mv{v}_k)^H(\mv{h}_k+\tau_t\mv{v}_k)}{M}\right] \label{eqn:appendix 7} \\
& = \frac{\beta_k^2}{\beta_k+\tau_t^2},
\end{align}
where (\ref{eqn:appendix 7}) is due to (\ref{eqn:detection function 1})
as consequence of Theorem \ref{theorem1}.

Similarly, as $M\rightarrow\infty$, we can show that $\Delta \upsilon_k(M)$ converges to
$\frac{\beta_k \tau_\infty^2}{\beta_k+\tau_\infty^2}$.

\end{appendix}


\begin{thebibliography}{1}
\bibitem{LiangISIT} L. Liu and W. Yu, ``Massive device connectivity with massive MIMO,'' in {\it Proc. IEEE Inter. Symp. Inf. Theory (ISIT)}, Jun. 2017.

\bibitem{Bockelmann} C. Bockelmann, N. Pratas, H. Nikopour, K. Au, T. Svensson, C. Stefanovic, P. Popovsk, and A. Dekorsy, ``Massive machine-type communications in 5G: Physical and MAC-layer solutions,'' {\it IEEE Commun. Mag.}, vol. 54, no. 9, pp. 59-65, Sep. 2016.

\bibitem{yu_ITA} W. Yu, ``On the fundamental limits of massive connectivity,`` in {\it Proc. Inf. Theory and Appl. (ITA) Workshop}, Feb. 2017.

\bibitem{donoho_amp}  D. L. Donoho, A. Maleki, and A. Montanari, ``Message-passing algorithms for compressed sensing,`` {\it Proc. Nat. Acad. Sci.}, vol. 106,
no. 45, pp. 18914-18918, Nov. 2009.

\bibitem{LiuPart2} L. Liu and W. Yu, ``Massive connectivity with massive MIMO-Part II: Achievable rate characterization,'' {\it IEEE Trans. Signal Process.}, 2018, to be published.

\bibitem{Niyato}  M. Hasan, E. Hossain, and D. Niyato, ``Random access for machine-to-machine communication in LTE-advanced networks: Issues
and approaches,`` {\it IEEE Commun. Mag.}, vol. 51, no. 6, pp. 86-93, Jun. 2011.

\bibitem{Pratas} N. K. Pratas, H. Thomsen, C. Stefanovic, and P. Popovski, ``Codeexpanded random access for machine-type communications,`` in {\it Proc. IEEE
Globecom Workshops}, Dec. 2012, pp. 1681-1686.

\bibitem{Bjornson}  E. Bj\"ornson, E. de Carvalho, J. H. S{\o}rensen, E. G. Larsson, and P. Popovski, ``A random access protocol for pilot allocation in crowded
massive MIMO systems,`` {\it IEEE Trans. Wireless Commun.}, vol. 16, no. 4, pp. 2220-2234, Apr. 2017..

\bibitem{zhu} H. Zhu and G. B. Giannakis, ``Exploiting sparse user activity in multiuser detection,`` {\it IEEE Trans. Commun.}, vol. 59, no. 2, pp. 454-465, Feb. 2011.

\bibitem{dekorsy1} H. F. Schepker and A. Dekorsy, ``Compressive sensing multi-user detection with block-wise orthogonal least squares,`` in {\it Proc. IEEE Veh.
Tech. Conf. (VTC Spring)}, May 2012, pp. 1-5.

\bibitem{xu_rao_lau}  X. Xu, X. Rao, and V. K. N. Lau, ``Active user detection and channel estimation in uplink CRAN systems,'' in {\it Proc. IEEE Int. Conf. Commun. (ICC)}, Jun. 2015, pp. 2727-2732.

\bibitem{wunder2}  G. Wunder, P. Jung, and M. Ramadan, ``Compressive random access using a common overloaded control channel,'' in {\it Proc. IEEE Int. Conf. Commun. (ICC) Workshops}, Jun. 2015.

\bibitem{hannak} G. Hannak, M. Mayer, A. Jung, G. Matz, and N. Goertz, ``Joint channel estimation and activity detection for multiuser communication
systems,'' in {\it Proc. IEEE Int. Conf. Commun. (ICC) Workshops}, Jun. 2015, pp. 2086-2091.

\bibitem{dekorsy2} H. F. Schepker, C. Bockelmann, and A. Dekorsy, ``Exploiting sparsity in channel and data estimation for sporadic multi-user
communication,`` in {\it Inter. Symp. Wireless Commun. Sys. (ISWCS)}, Aug. 2013, pp. 1-5.

\bibitem{wunder_5G} G. Wunder, H. Boche, T. Strohmer, and P. Jung, ``Sparse signal processing concepts for efficient 5G system design,`` {\it IEEE Access},
vol. 3, pp. 195-208, 2015.

\bibitem{wunder1} G. Wunder, P. Jung, and C. Wang, ``Compressive random access for post-LTE systems,`` in {\it Proc. IEEE Int. Conf. Commun. (ICC) Workshops},
Jun. 2014.

\bibitem{zhilin_ICASSP} Z. Chen and W. Yu, ``Massive device activity detection by approximate message passing,`` in {\it Proc. IEEE Int. Conf. Acoustic, Speech, Signal
Process. (ICASSP)}, Mar. 2017.

\bibitem{zhilin_JSAC} Z. Chen, F. Sohrabi, and W. Yu, ``Sparse activity detection for massive connectivity,`` {\it IEEE Trans. Signal Process.}, vol. 66, no. 7, pp. 1890-1904, Apr. 2018.

\bibitem{bayati_montanari_amp}  M. Bayati and A. Montanari, ``The dynamics of message passing on dense graphs, with applications to compressed sensing,`` {\it IEEE
Trans. Inf. Theory}, vol. 57, no. 2, pp. 764-785, Feb. 2011.

\bibitem{marzetta} T. L. Marzetta, ``Noncooperative cellular wireless with unlimited numbers of base station antennas,`` {\it IEEE Trans. Wireless Commun.},
vol. 9, no. 11, pp. 3590-3660, Nov. 2010.

\bibitem{larsson} H. Q. Ngo, E. G. Larsson, and T. L. Marzetta, ``Energy and spectral efficiency of very large multiuser MIMO systems,`` {\it IEEE Trans.
Commun.}, vol. 61, no. 4, pp. 1436-1449, Apr. 2013.

\bibitem{debbah}  J. Hoydis, S. Brink, and M. Debbah, ``Massive MIMO in the UL/DL of cellular networks: How many antennas do we need?`` {\it IEEE
J. Sel. Areas Commun.}, vol. 31, no. 2, pp. 160-171, Feb. 2013.

\bibitem{larsson14}  E. G. Larsson, F. Tufvesson, O. Edfors, and T. L. Marzetta, ``Massive MIMO for next generation wireless systems,`` {\it IEEE Commun.
Mag.}, vol. 52, no. 2, pp. 186-195, Feb. 2014.

\bibitem{debbah12} S. Wagner, R. Couillet, M. Debbah, and D. T. M. Slock, ``Large system analysis of linear precoding in correlated MISO broadcast
channels under limited feedback,`` {\it IEEE Trans. Inf. Theory}, vol. 58, no. 7, pp. 4509-4537, Jul. 2012.

\bibitem{guo3} X. Chen, T.-Y. Chen, and D. Guo, ``Capacity of Gauassian many-access channels,`` {\it IEEE Trans. Inf. Theory}, vol. 63, no. 6, pp.
3516-3519, Jun. 2017.

\bibitem{baron_MMV}  J. Kim, W. Chang, B. Jung, D. Baron, and J. C. Ye, ``Belief propagation for joint sparse recovery,`` Feb. 2011, [Online] Available:
http://arxiv.org/abs/1102.3289.

\bibitem{schniter_mmv}  J. Ziniel and P. Schniter, ``Efficient high-dimensional inference in the multiple measurement vector problem,`` {\it IEEE Trans. Signal
Process.}, vol. 61, no. 2, pp. 340-354, Jan. 2013.

\bibitem{rangan3}  S. Rangan, ``Generalized approximate message passing for estimation with random linear mixing,`` in {\it Proc. IEEE Int. Symp. Inf. Theory
(ISIT)}, Jul. 2011, pp. 2168-2172.

\bibitem{Gautschi} W. Gautschi, ``The incomplete gamma functions since Tricomi,`` {\it Atti dei Convegni Linci}, no. 1998, pp. 203-237, 2011.


\end{thebibliography}
\end{document}